\documentclass
[aps,prb,twocolumn,floatfix,english,showpacs,10pt,superscriptaddress]{revtex4-2}
\usepackage{graphicx}
\usepackage{booktabs}
\usepackage{amsmath}
\usepackage{physics}
\usepackage{amssymb}
\usepackage{colordvi}
\usepackage{verbatim}
\usepackage{xcolor}
\usepackage{mathrsfs}
\usepackage{epsfig}
\usepackage{lipsum}
\usepackage{amsfonts}
\usepackage{makecell}

\usepackage[unicode=true, breaklinks=false, pdfborder={0 0 1}, backref=false,
colorlinks=true, linkcolor=blue, urlcolor=blue, citecolor=blue]{hyperref}%
\newcommand{\bgreek}[1]{\mbox{\boldmath$#1$\unboldmath}}

\providecommand{\U}[1]{\protect\rule{.1in}{.1in}}
\setcitestyle{numbers,square}

\begin{document}

\title{Directional entanglement of spin-orbit locked nitrogen-vacancy centers by magnons}

\author{Zhiping Xue}

\affiliation{School of Physics, Huazhong University of Science and Technology, Wuhan 430074, China}

\author{Ji Zou}
\affiliation{Department of Physics, University of Basel, Klingelbergstrasse 82, 4056 Basel, Switzerland}

\author{Chengyuan Cai}

\affiliation{School of Physics, Huazhong University of Science and Technology, Wuhan 430074, China}

\author{Gerrit E. W. Bauer}

\affiliation{Kavli Institute for Theoretical Sciences, University of the Chinese Academy of Sciences, Beijing 100190, China}
\affiliation{WPI-AIMR and Institute for Materials Research and CSIS, Tohoku University, Sendai 980-8577, Japan}

\author{Tao Yu}
\email{taoyuphy@hust.edu.cn}
\affiliation{School of Physics, Huazhong University of Science and Technology, Wuhan 430074, China}

\date{\today}

\begin{abstract}

We address that the stray magnetic field emitted by the excited quantum states of the nitrogen-vacancy (NV) centers is spin-momentum locked, such that the spin transfer to nearby ferromagnetic nanostructures is unidirectional. This may allow the controlled excitation of propagating magnons by NV centers in diamond. A pair of NV spin qubits exchange virtual magnons in a magnetic nanowire in a chiral manner that leads to directional quantum entanglement. A magnon-based ``quantum-entanglement isolator" should be a useful device in future quantum information technology.

\end{abstract}

\maketitle

\section{Introduction}

Nitrogen-vacancy (NV) centers in diamond~\cite{gruber1997scanning} are interesting for quantum information processing, e.g., in optical networks~\cite{BrouriOL,Ruf,Bradley} and as the source for single photons~\cite{Schroder,Riedelprx,Panadero} and single surface plasmons~\cite{single_plasmon}. As ultrasensitive magnetic field detectors, they have shown their worth in measuring spin relaxation times~\cite{HansonPRL,Kennedy} and spin-coherent dynamics~\cite{HansonScience,WaldherrPRL}. Room-temperature nanoscale magnetometry~\cite{GaebelNP,CasolaNRM}, detection of the Meissner effect~\cite{BouchardNJP,Lesik,YipSci}, and pressure-driven phase transitions~\cite{Hsiehpre} have been other feats. Single or few NV centers are created by controlled nitrogen ion implantation into diamond~\cite{CoherentGQ,GaebelNP,MazeNature,Hong,Maletinsky,CooperPRL,Thiel,Gieseler,Conangla,Schirhagl,FincoNC,Huxter}. In the vicinity of magnetic insulators, e.g., yttrium iron garnet (YIG) and van der Waals two-dimensional magnets, NV centers detect the stray magnetic fields of spin waves ~\cite{CoherentGQ,Andrich,WolfePRB,WolfeAPL,Sar,ChunhuiDu,Iacopo,Borstsci}, paving the way towards exploring magnetic excitation that may be highly localized~\cite{GuoPRL} or in the non-linear regime~\cite{Mehrdad_nonlinear}.  The magnetic noise generates the fluctuation of the magnetic stray field that chirally couples with the NV centers, allowing for their sensing via quantum impurity relaxometry~\cite{chiral_noise}.
In order to serve as qubits in a scalable platform for quantum computation, NV centers must be coherently coupled. Since the direct coupling between two NVs is negligibly small when separated by more than 30~nm~\cite{Dolde}, an indirect coupling by real and virtual magnon exchange with proximity magnets becomes attractive~\cite{FukamiPRX,Xin-LeiHei,Trifunovic,Candido,FukamiPRX,Toyli}. References~\cite{FukamiPRX,Toyli,Candido,Toeno} showed that the coupling is coherent and reciprocal. 

\begin{figure}[htb!]
\centering\includegraphics[width=0.46\textwidth,trim=0.6cm 0cm 0cm 0.1cm]{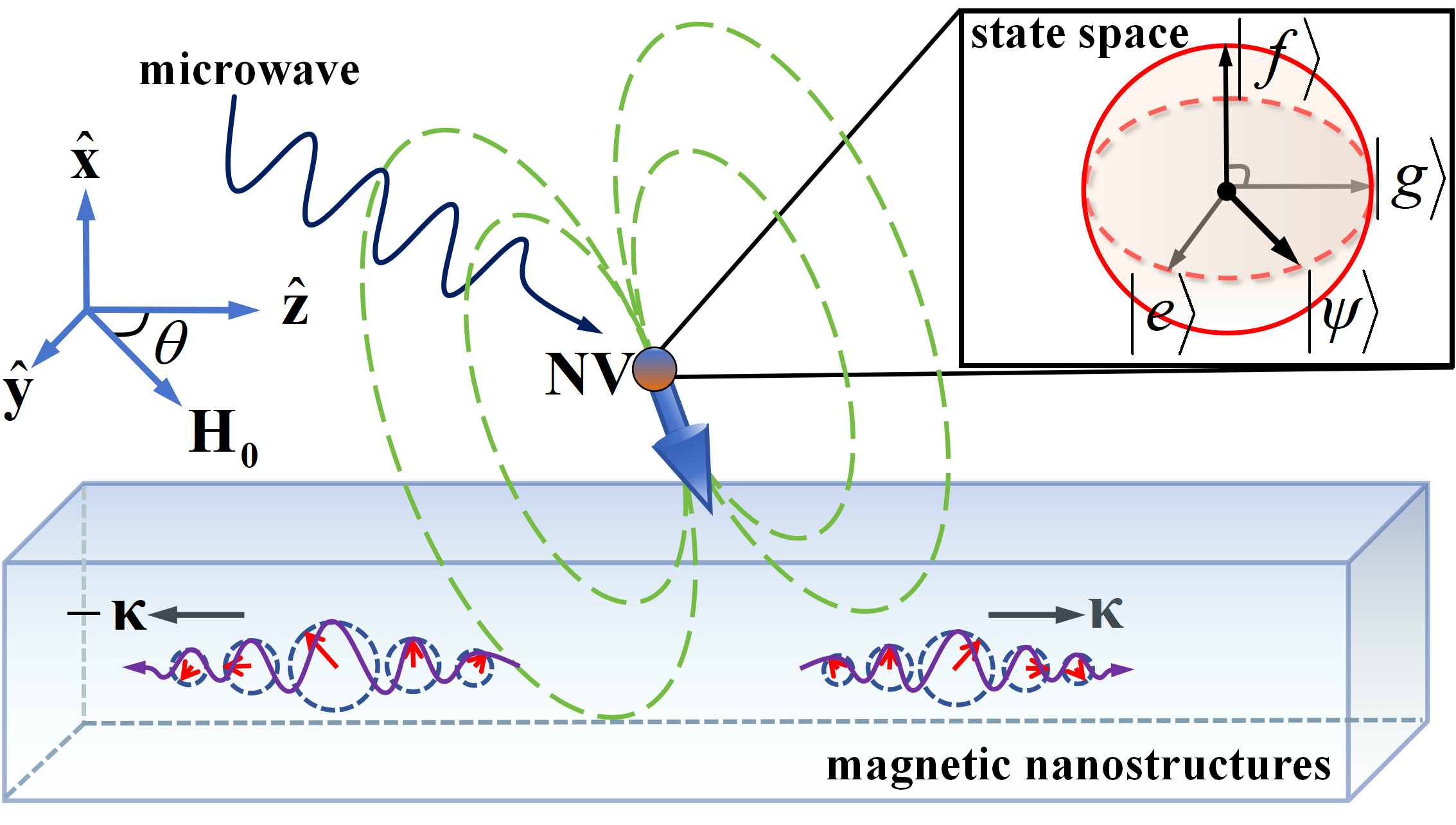} 
\caption{Schematic of NV centers in diamonds placed on top of a magnetic nanostructure, biased by the applied magnetic field $\mathbf{H}_{0}$ with an angle $\theta$ to the $\hat{\bf z}$-direction. The uniaxial anisotropy direction $\hat{\bf n}_{\rm NV}$ of the NV center is set to be along the $\hat{\bf z}$-direction. \textcolor{blue}{The thick downward arrow leaving the NV center represents its time-dependent magnetic moment. } The inset sketches the quantum state $|\psi\rangle$ of the NV center spanned by the ground $|g\rangle$ and excited $|e\rangle$ states, \textcolor{blue}{with $\{|g\rangle,|e\rangle, |f\rangle\}$ denoting, respectively, the ground, the first excited, and the second excited eigenstate of the NV spin-one operator in the Dirac notation.} }
\label{NV_center_1}
\end{figure}
 
In this work, we focus on the chirality of the stray magnetic fields emitted by an NV center in diamond. We predict that the conservation of angular momentum renders the spin transfer to magnons in ferromagnetic nanostructures excited by NV centers to be unidirectional. Nakamura \textit{c.s.}~\cite{Science_1,Science_2} and You \textit{c.s.}~\cite{PRL_single_magnon} demonstrated the creation and manipulation of single magnons by confining a YIG sphere and a superconducting qubit to a microwave cavity. There, only magnons in the uniformly precessing (Kittel) mode couple to the cavity microwaves. However, future quantum magnonic circuits demand controlled excitation of single \textit{propagating} magnons~\cite{roadmap_2024}. Analogous to the photostable single-photon~\cite{Schroder,Riedelprx,Panadero} or surface plasmon~\cite{single_plasmon} generation, the NV-center allows excitation of  propagating magnons as illustrated in Fig.~\ref{NV_center_1} for an NV center on top of a magnetic nanostructure with an in-plane magnetic field ${\bf H}_{0}$ applied at an angle $\theta$ with the $\hat{\mathbf{z}}$-axis. 
Fukami \textit{et al.} reported a strong indirect coupling between two NV centers via a magnetic wire~\cite{FukamiPRX} but limited themselves to the reciprocal coupling for saturation magnetizations aligned with the wire direction. Here, we focus on the chiral/nonreciprocal coupling that emerges when magnetizations are normal to the wire and reveal an enhanced quantum entanglement in the dispersive regime in which the NV center frequencies lie in the magnon \textcolor{blue}{band}. This chiral interaction between NV spin qubits leads to the nonreciprocal formation of quantum entanglement, i.e., depending on which qubit is initially in its excited state, which promises the functionality of an isolator for the quantum entanglement in the quantum circuits.

This article is organized as follows. In Sec.~\ref{dipolar_chirality}, we build up the quantum dynamics of the NV-center excitations and demonstrate the chirality of their dipolar stray fields. We show that such dipolar chirality leads to the chiral coupling between magnons and NV spin qubits in 
Sec.~\ref{magnetic_nanostructure}. In Sec.~\ref{nonreciprocal_entanglement}, we show that the photonic spin-orbit coupling leads to the nonreciprocal quantum entanglement
between two NV centers, which may act as a quantum-entanglement isolator. We summarize our results and give an outlook in Sec.~\ref{summary}.

\section{Dipolar chirality of NV spin excitations}
\label{dipolar_chirality}

The dynamics of an NV center in diamonds are governed by the Hamiltonian 
\begin{align}
\hat{H}_{\mathrm{NV}}=\textcolor{blue}{D_{\mathrm{NV}}(\hat{\bf n}_{\mathrm{NV}}\cdot\hat{\mathbf{S}}_{\mathrm{NV}})^2/\hbar}+\mu_0\gamma\hat{\mathbf{S}}_{\mathrm{NV}}\cdot\mathbf{H}_{0},
\end{align}
where $D_{\mathrm{NV}}=2\pi\times2.877~\mathrm{GHz}$ is the zero-field splitting~\cite{Iacopo}, $\gamma=2\pi\times 28 ~\mathrm{GHz}/\mathrm{T}$ is the electron gyromagnetic ratio, $\mu_0$ is the vacuum permeability, and $\hat{\bf S}_{\rm NV}$ is the spin-one operator of the NV center. The uniaxial anisotropy direction $\hat{\bf n}_{\rm NV}$ of the NV center is set to be along the in-plane $\hat{\bf z}$-direction as in Fig.~\ref{NV_center_1}. Since $|{\bf S}_{\rm NV}|=\hbar$ with eigenstates $|1,m\rangle$ with $m=\{-1,0,1\}$ being the azimuthal quantum number along the $\hat{\bf z}$-direction, we expand the spin operator in the Hilbert space spanned by the states $\{|1,1\rangle, |1,0\rangle,|1,-1\rangle\}$ as 
\begin{align}
   \hat{\mathbf{S}}_{\rm NV}
   &=\sum_{m,n=\{-1,0,1\}}|1,m\rangle\langle 1,n| \langle 1,m|\hat{\mathbf{S}}_{\rm NV}|1,n\rangle\nonumber\\
   &=\sum_{m,n=\{-1,0,1\}}|1,m\rangle\langle 1,n|\bgreek{\cal S}_{mn}, 
   \label{spin_operators}
\end{align}
where
\begin{align}
&{\mathbf{\cal{S}}}_{\mathrm{NV}}^x=\frac{\hbar}{\sqrt{2}}\left(\begin{matrix}
    0 & 1  & 0   \\
    1& 0 & 1   \\
    0 & 1 & 0 
      \end{matrix}\right),  
    ~~~{\mathbf{\cal{S}}}_{\mathrm{NV}}^y=\frac{\hbar}{\sqrt{2}}\left(\begin{matrix}
    0 & -i  & 0   \\
    i& 0 & -i  \\
    0 & i & 0 
    \end{matrix}\right),\nonumber\\
    &{{\cal S}}_{\mathrm{NV}}^z=\hbar\left(\begin{matrix}
    1 & 0  & 0   \\
    0& 0 & 0   \\
    0 & 0 & -1 
    \end{matrix}\right).
\end{align}
The Hamiltonian matrix for the NV center then reads 
\[
\cal{H}_{\rm{NV}}=\hbar\left(\begin{matrix}
      D_{\mathrm{NV}}+\omega_{\rm H}\cos\theta & -\frac{i}{\sqrt{2}}\omega_{\rm H}\sin\theta  & 0   \\
      \frac{i}{\sqrt{2}}\omega_{\rm H}\sin\theta & 0 & -\frac{i}{\sqrt{2}}\omega_{\rm H}\sin\theta   \\
      0 & \frac{i}{\sqrt{2}}\omega_{\rm H}\sin\theta & D_{\mathrm{NV}}-\omega_{\rm H}\cos\theta
      \end{matrix}\right),
      \]
where $\omega_{\rm H}=\mu_0\gamma H_0$.

 $\{\hbar\omega_f, \hbar\omega_e, \hbar\omega_g\}$ are the three eigenvalues of $\cal{H}_{\rm{NV}}$ in the decreasing order. The corresponding eigenstates are $\{u_f, u_e, u_g\}$ and unitary transformation matrix $U=(u_f,u_e,u_g$) such that $(|1,1\rangle, |1,0\rangle,|1,-1\rangle)=(|f\rangle,|e\rangle,|g\rangle)U^{\dagger}$ diagonalizes the Hamiltonian 
\begin{align}
\hbar\left(\begin{matrix}
      \omega_f & 0  & 0   \\
      0& \omega_e & 0   \\
      0 & 0 & \omega_g 
\end{matrix}\right)=U^{\dagger}{\cal{H}_{\rm{NV}}}U.
\label{diagonalization}
\end{align} 
\textcolor{blue}{We note $\{|f\rangle,|e\rangle,|g\rangle\}$ are the basis of the unperturbed Hamiltonian of the NV centers in the Dirac notation, while $u_f$, $u_e$, and $u_g$ are three column vectors of mode amplitudes.}
When $\theta=0$, ${\bf H}_0\parallel \hat{\bf n}_{\rm NV}\parallel \hat{\bf z}$,  $\{\omega_f,\omega_e,\omega_g\}=\{D_{\rm NV}+\mu_0\gamma H_0,D_{\rm NV}-\mu_0\gamma  H_0,0\}$, and the eigenvectors $u_f=(1,0,0)^T$, $u_e=(0,0,1)^T$, and $u_g=(0,1,0)^T$, where we assume $D_{\rm NV}>\mu_0\gamma  H_0$ or $\mu_0H_0<0.1$~T. When $\mu_0H_0>0.1$~T or $D_{\rm NV}<\mu_0\gamma  H_0$,  $\{\omega_f,\omega_e,\omega_g\}=\{D_{\rm NV}+\mu_0\gamma H_0,0,D_{\rm NV}-\mu_0\gamma  H_0\}$.

The emitted stray fields of the excited NV centers can be controlled by the static (in-plane) magnetic-field directions $\theta$ as in Fig.~\ref{NV_center_1}. \textcolor{blue}{Upon applying a weak polarized microwave field} $h_0\sin(\omega_{\rm ext}t)\hat{\bf y}$ with amplitude $h_0$ and frequency $\omega_{\rm ext}$, the Hamiltonian acquires a time-dependent interaction term 
\begin{align}
 \hat{H}&=\hbar \omega_{f}|f\rangle \langle f|+\hbar \omega_{e}|e\rangle \langle e|+\hbar \omega_{g}|g\rangle \langle g|\nonumber\\
    &+\hbar\sin(\omega_{{\rm ext}}t)(|f\rangle,|e\rangle,|g\rangle){\cal G}\left(\begin{matrix}
     \langle f|  \\
     \langle e| \\
    \langle g|
    \end{matrix}\right),
\end{align}
where ${\cal G}=\mu_0\gamma h_0 U^{\dagger}{\mathbf{\cal{S}}}_{\mathrm{NV}}^yU/\hbar$. 
\textcolor{blue}{Here,  we focus on the ``photon spin" carried by the dipolar field emitted by an excited NV center that can be revealed by its quantum state $|\psi(t) \rangle$ during its initial time evolution at which interaction with the environment does not yet kick in. We thereby disregard the interaction with the environment when solving a Schr\"odinger equation. This approximation is allowed for times sufficiently shorter than the dephasing time of the NV center.  In the typical case with $\theta=0.4\pi$ and $\mu_0H_0=0.05 $~T,  $\omega_{\rm NV}\equiv\omega_e-\omega_g\approx20$~GHz, so the period of evolution $T_{\rm NV}={2\pi}/{\omega_{\rm NV}}\approx0.3~{\rm ns}$ is much smaller than the dephasing time of an NV center, i.e., $\sim 1~{\rm ms}$~\cite{Herbschleb}.}

\begin{figure}[htb]
\centering
\includegraphics[width=0.48\textwidth,trim=0.6cm 0cm 0cm 0.1cm]{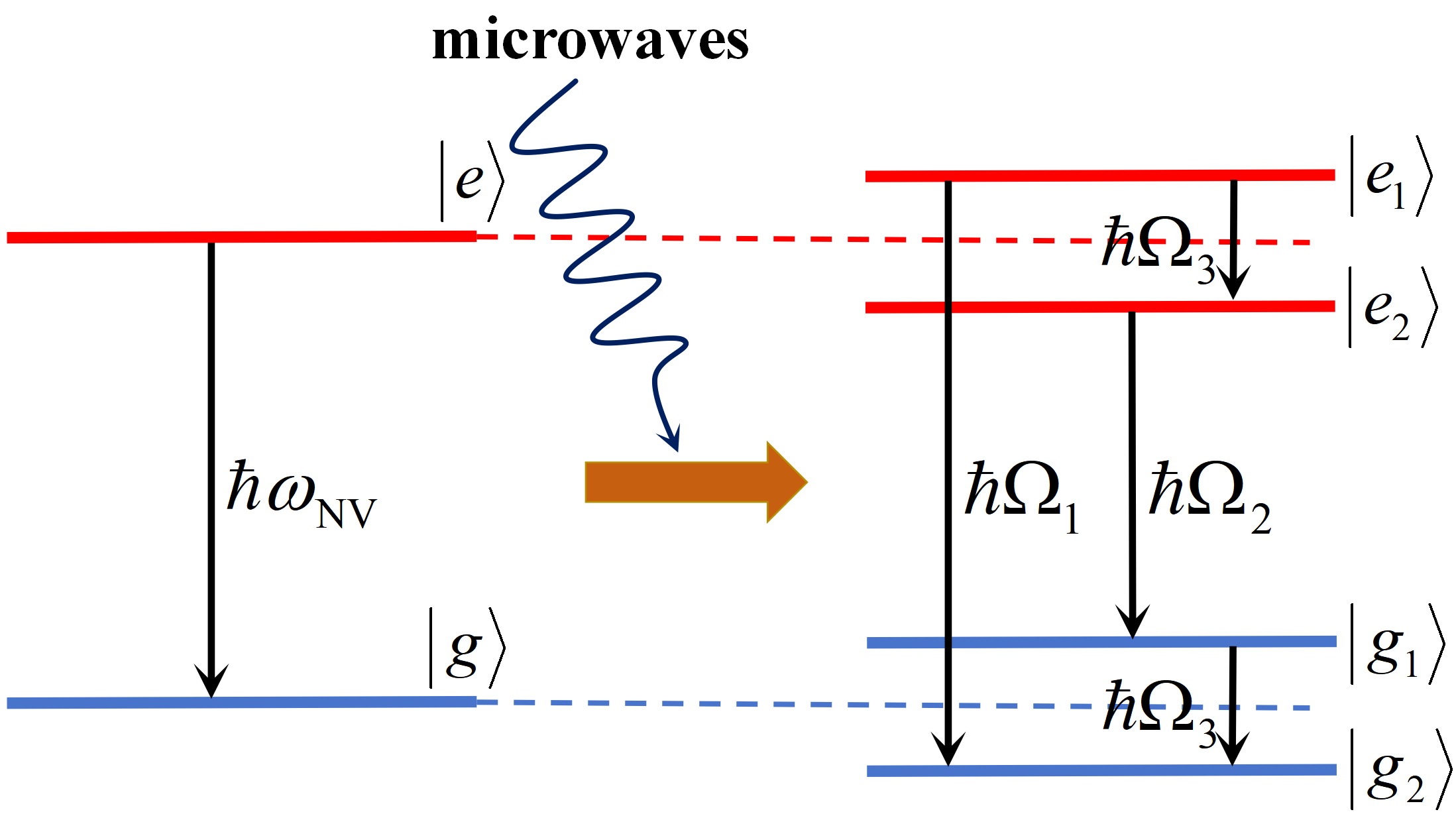}
    \caption{Photo-induced splitting of the NV center eigenfrequency by a resonant microwave drive $h_0\sin(\omega_{\rm NV}t)\hat{\bf y}$. Here $\Omega_1=\omega_{\rm NV}+|G_{eg}|$, $\Omega_2=\omega_{\rm NV}-|G_{eg}|$, and $\Omega_3=|G_{eg}|$, where \(\omega_{\rm NV}\) is the bare resonance frequency and \(G_{eg}\) is proportional to the amplitude of the driving field. }
    \label{energy_level_splitting}
\end{figure}

The wave function of the NV center carries out Rabi-like oscillations (refer to Appendix~\ref{photon_spin})
\begin{align}
    |\psi(t)\rangle&=\frac{1}{2}\left(e^{-i(\omega_g+|G_{eg}|/2)t}+e^{-i(\omega_g-|G_{eg}|/2)t}\right)|g\rangle\nonumber\\
    &-\frac{i}{2}\left(e^{-i(\omega_e-|G_{eg}|/2)t}-e^{-i(\omega_e+|G_{eg}|/2)t}\right)|e\rangle,
     \label{general_solution}
\end{align}
in which the microwave field couples the ground $|g\rangle$ and excited $|e\rangle$ states by
\[
G_{eg}=G_{ge}^*=\mu_0\gamma h_0u_e^{\dagger}{\mathbf{\cal{S}}}_{\mathrm{NV}}^yu_g/\hbar. 
\]
The populations of the excited $|C_e(t)|^2$ and ground  $|C_g(t)|^2$ states oscillate with eigenfrequencies that are blue and red-shifted by $|G_{eg}|$. When ${\bf H}_0\parallel \hat{\bf z}$, i.e., $\theta=0$, $|G_{eg}|={\mu_0\gamma h_0}/\sqrt{2}$. The four frequencies of the driven NV center
\begin{align}
    \omega_1&=\omega_g+{|G_{eg}|}/{2},\nonumber\\
    \omega_2&=\omega_g-{|G_{eg}|}/{2},\nonumber\\
    \omega_3&=\omega_e+{|G_{eg}|}/{2},\nonumber\\
    \omega_4&=\omega_e-{|G_{eg}|}/{2},
    \label{quasi_energies}
\end{align}
plotted in Figure~\ref{energy_level_splitting} show a field-induced splitting of the states $|g\rangle$ and $|e\rangle$.

\textcolor{blue}{The $j$-component of the spin operator Eq.~\eqref{spin_operators} reads
\begin{align}
    \hat{S}_{\rm NV}^j
   &=\sum_{m,n=\{-1,0,1\}}|1,m\rangle\langle 1,n|\bgreek{\cal S}^j_{mn}\nonumber\\
    &=(|f\rangle, |e\rangle,|g\rangle)\left(\begin{array}{ccc}
       S^j_{ff}  & S^j_{fe} & S^j_{fg} \\
       S^j_{ef} & S^j_{ee} & S^j_{eg}\\ S^j_{gf}  & S^j_{ge} & S^j_{gg}
    \end{array}\right)\left(\begin{array}{c}
         \langle f | \\
         \langle e|  \\
         \langle g|
    \end{array}\right),
\end{align} 
where via the unitary transformation matrix $U=(u_f,u_e,u_g)$, $(|1,1\rangle, |1,0\rangle,|1,-1\rangle)U=(|f\rangle, |e\rangle,|g\rangle)$ and 
\begin{align}
    \left(\begin{array}{ccc}
       S^j_{ff}  & S^j_{fe} & S^j_{fg} \\
       S^j_{ef} & S^j_{ee} & S^j_{eg}\\ S^j_{gf}  & S^j_{ge} & S^j_{gg}
\end{array}\right)&=U^{\dagger}\left(\begin{array}{ccc}
       {\cal S}^j_{1,1}  & {\cal S}^j_{1,0} & {\cal S}^j_{1,-1} \\
       {\cal S}^j_{0,1} & {\cal S}^j_{0,0} & {\cal S}^j_{0,-1}\\ {\cal S}^j_{-1,1}  & {\cal S}^j_{-1,0} & {\cal S}^j_{-1,-1}
    \end{array}\right)U.
\end{align}
In terms of the eigenstates $\{|g\rangle,|e\rangle,|f\rangle\}$, $S_{\alpha\beta}^j=\langle \alpha|\hat{S}_{\rm NV}^j|\beta\rangle$ are the matrix elements of the spin operator. When  mixing with the $|f\rangle$ state may be disregarded, we may focus on the submatrix spanned by the ground $|g\rangle$ and excited $ |e\rangle$ states}
\begin{align}
    \hat{S}_{\rm NV}^j&=(|e\rangle,|g\rangle)\left(\begin{array}{cc}
        S^j_{ee} & S^j_{eg}\\  S^j_{ge} & S^j_{gg}\end{array}\right)\left(\begin{array}{c}
         \langle e|  \\
         \langle g|
\end{array}\right)\nonumber\\&=\sum_{\alpha,\beta\in \{e,g\}}\langle \alpha|\hat{S}_{\rm NV}^j|\beta\rangle|\alpha\rangle\langle \beta|\nonumber\\
&=S_{eg}^{j}\hat{\sigma}^{+}+S_{ge}^{j}\hat{\sigma}^{-}+S_{gg}^{j}\hat{\sigma}^{-}\hat{\sigma}^{+}+S_{ee}^{j}\hat{\sigma}^{+}\hat{\sigma}^{-},
    \label{Ssigma}
\end{align}
where $\hat{\sigma}^{-}=|g\rangle\langle e|$ and $\hat{\sigma}^{+}=|e \rangle \langle g|$ represent the annihilation and creation operators in the spin-1 Hilbert space, and the matrix elements read $S_{eg}^{\beta}=\langle e|\hat{S}_{\rm NV}^{\beta}|g\rangle=u_{e}^{\dagger}{\cal S}_{\rm NV}^{\beta}u_{g}$, $S_{ge}^{\beta}=u_{g}^{\dagger}{\cal S}_{\rm NV}^{\beta}u_{e}=S_{eg}^{\beta*}$, $ S^{\beta}_{gg}=u_{g}^{\dagger}{\cal S}_{\rm NV}^{\beta}u_{g}$, and $S_{ee}^{\beta}=u_{e}^{\dagger}{\cal S}_{\rm NV}^{\beta}u_{e}$.

The spin of an NV center corresponds to a magnetic moment  
\[
\hat{m}_\beta=-\gamma \hat{S}^{\beta}_{\rm NV}\delta({\bf r})
\]
that generates the magnetic stray field 
\begin{align}
     \hat{H}_{\beta}(\mathbf{r},t)=\frac{1}{4\pi}\partial_{\beta}\partial_{\alpha
}\int_{-\infty}^{\infty}d{\mathbf{r}}^{\prime}\frac{\hat{m}_{\alpha}(\mathbf{r^{\prime}},t)}{|\mathbf{r}-\mathbf{r^{\prime}}|},
\end{align}
\textcolor{blue}{in which the Einstein summation is adopted for repeated $\alpha=\{x,y,z\}$ throughout.}
For $x<0$, \textit{i.e.}, under the NV center,    
\begin{align}
\hat{H}_{x}({\bf r},t)&=\sum_{q_y,q_z}\hat{\mathcal{F}}_{q_y,q_z}\frac{1}{2}e^{i(q_yy+q_zz)}e^{\sqrt{q_y^2+q_z^2}x}+{\rm H.c.},\nonumber\\
\hat{H}_{y}({\bf r},t)&=\sum_{q_y,q_z}\hat{\mathcal{F}}_{q_y,q_z}\frac{iq_y}{2\sqrt{q_y^2+q_z^2}}e^{i(q_yy+q_zz)}e^{\sqrt{q_y^2+q_z^2}x}+{\rm H.c.},\nonumber\\
\hat{H}_{z}({\bf r},t)&=\sum_{q_y,q_z}\hat{\mathcal{F}}_{q_y,q_z}\frac{iq_z}{2\sqrt{q_z^2+q_y^2}}e^{i(q_yy+q_zz)}e^{\sqrt{q_y^2+q_z^2}x}+{\rm H.c.},
\label{stray_field_NV_operator}
\end{align}
where $\hat{\mathcal{F}}_{q_y,q_z}=-\gamma\left(\sqrt{q_y^2+q_z^2}\hat{S}^x_{\rm NV}+iq_y\hat{S}^y_{\rm NV}+iq_z\hat{S}^z_{\rm NV}\right)$ is a form factor.
The expectation value of the dipolar magnetic field oscillates with frequencies  $\Omega_{n=1,2,3}$ 
\begin{align}
H_{\beta}(\mathbf{r},t)&=\langle \psi(t)|\hat{H}_{\beta}(\mathbf{r},t)|\psi(t)\rangle \nonumber\\
&=\sum_{n=\{1,2,3\}}\sum_{q_{y},q_{z}}e^{i(q_{y}y+q_{z}z)-i\Omega_n t}H^{(n)}_{\beta}(x;q_{y},q_{z}
)\nonumber\\
&+{\rm H.c.}, 
\label{dipolar_field1}
\end{align}
where the Fourier components of the dipolar field Eq.~(\ref{dipolar_field1}) read (for $x<0$)
\begin{align}
H^{(n)}_{x}(x,q_{y},q_{z}
) &  =\mathcal{F}^{(n)}_{q_{y},q_{z}} e^{\sqrt{q_{y}^{2}+q_{z}^{2}}x}\frac{1}{2},\nonumber\\
H^{(n)}_{y}(x,q_{y},q_{z}
) &=\mathcal{F}^{(n)}_{q_{y},q_{z}}e^{\sqrt{q_{y}^{2}+q_{z}^{2}}x}\frac{iq_y}{2\sqrt{q_{y}^{2}+q_{z}^{2}}},\nonumber\\
H^{(n)}_{z}(x,q_{y},q_{z}
) &=\mathcal{F}^{(n)}_{q_{y},q_{z}} e^{\sqrt{q_{y}^{2}+q_{z}^{2}}x}\frac{iq_z}{2{\sqrt{q_{y}^{2}+q_{z}^{2}}}},
\label{field_small}
\end{align}
with form factors 
\begin{align}
    \mathcal{F}^{(1)}_{q_y,q_z}&=-\gamma\frac{i}{2}\left(S^x_{ge}\sqrt{q_y^2+q_z^2}+iS^y_{ge}q_y+iS^z_{ge}q_z\right),\nonumber\\
\mathcal{F}^{(2)}_{q_y,q_z}&=\gamma\frac{i}{2}\left(S^x_{ge}\sqrt{q_y^2+q_z^2}+iS^y_{ge}q_y+iS^z_{ge}q_z\right),\nonumber\\
\mathcal{F}^{(3)}_{q_y,q_z}&=-\gamma\frac{1}{2}\Big[(S^x_{gg}-S^x_{ee})\sqrt{q_y^2+q_z^2}\nonumber\\
&+i\left(S^y_{gg}-S^y_{ee}\right)q_y+i\left(S^z_{gg}-S^z_{ee}\right)q_z\Big].
\end{align}
When $|q_y|\ll |q_z|$, the momentum \(\mathbf{q}\) of the stray field is locked to its circular polarization: $|H^{(n)}_y|\ll |H^{(n)}_{x,z}|$ and $H^{(n)}_z(x,q_y,q_z)\rightarrow i{\rm sgn}(q_z)H^{(n)}_x(x,q_y,q_z)$, such that the field is right or left circularly polarized depending the propagation direction $q_z>0$ or $q_z<0$, with spin axis along $\hat{\bf y}$.
When $|q_y|\gg|q_z|$, $|H^{(n)}_{x,y}(x,q_y,q_z)|\gg|H^{(n)}_z(x,q_y,q_z)|$ and $H^{(n)}_y(x,q_y,q_z)\rightarrow i{\rm sgn}(q_y)H^{(n)}_x(x,q_y,q_z)$, the circular polarization and momentum are still locked but the spin axis is along the $\hat{\bf z}$-direction.
When ${\bf H}_0\parallel \hat{\bf n}_{\rm NV}\parallel \hat{\bf z}$, the form factors simplify to $\mathcal{F}^{(1)}_{q_{y},q_{z}}=-i\gamma\hbar(\sqrt{q_y^2+q_z^2}+q_y)/(2\sqrt{2})$, $\mathcal{F}^{(2)}_{q_{y},q_{z}}=
i\gamma\hbar(\sqrt{q_y^2+q_z^2}+q_y)/(2\sqrt{2})$, and $\mathcal{F}^{(3)}_{q_{y},q_{z}}=
-i\gamma\hbar q_z /2$.

The evanescence of the emitted stray magnetic field \eqref{dipolar_field1} complicates the real-space interpretation of chirality in momentum space. 
Focusing therefore on a Fourier component with plane wave vector ${\bf q}$ it is convenient to introduce the photon spin density~\cite{Walker11,
Damon11, Kino1, Viktorov1,Bliokh1,chirality}
\begin{align}
    \mathbf{S}^{(n)}(x,{q_{y}},q_{z})=\frac{\mu_0}{4\Omega_i}\mathrm{Im} \left(  \mathbf{H}^{(n)\ast}
(x,q_{y},q_{z})\times\mathbf{H}^{(n)}(x,q_{y},q_{z})\right).
\end{align}
In the lower half-space ($x<0$) 
\begin{align}
S^{(n)}_{x}(x,{q_{y}},q_{z}) & =0,\nonumber\\
S^{(n)}_{y}(x,{q_{y}},q_{z}) & = -\frac{\mu_0}{8\Omega_i}|\mathcal{F}^{(n)}_{q_{y},q_{z}}|^{2} \frac{e^{2\sqrt{q_{y}^{2}+q_{z}^{2}}x}}{\sqrt{q_{y}%
^{2}+q_{z}^{2}}}q_{z},\nonumber\\
S^{(n)}_{z}(x,{q_{y}},q_{z}) & =\frac{\mu_0}{8\Omega_i}|\mathcal{F}^{(n)}_{q_{y},q_{z}}|^{2} \frac{e^{2\sqrt{q_{y}^{2}+q_{z}^{2}}x}}{\sqrt{q_{y}%
^{2}+q_{z}^{2}}}q_{y}.
\label{spin}
\end{align}
The spin density is purely transverse ${\bf q}\cdot {\bf S}^{(n)}_{\bf q}=0$ and right-handed with  chirality index $(-\hat{\bf x})\cdot(\hat{\bf S}^{(n)}_{\bf q}\times\hat{\bf q})=1$, which implies locking between the surface normal direction $-\hat{\bf x}$, the wave-vector direction $\hat{\bf q}$, and the spin direction $\hat{\bf S}_{\bf q}^{(n)}$.

\begin{figure}[tbh]
\centering
\includegraphics[width=0.48\textwidth,trim=0.6cm 0cm 0cm 0.1cm]{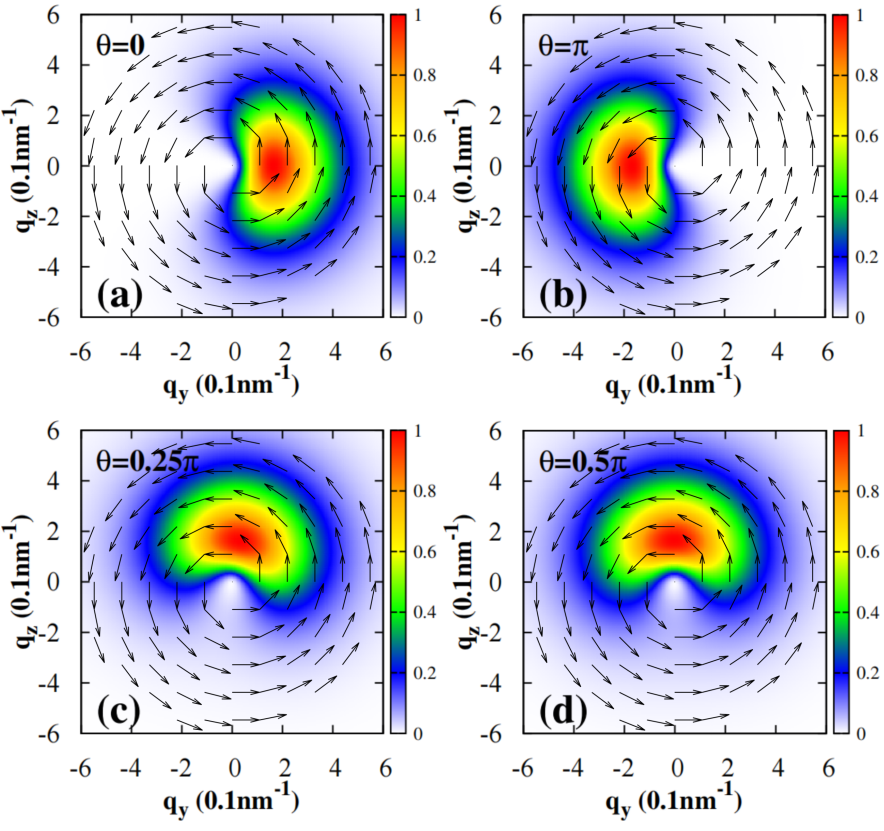}
\caption{Photon spin density $\mathbf{S}_{\bf q}^{(n=1)}(x=-6~\mathrm{nm})$ of the in-plane wave components \textbf{q} of the stray magnetic field emitted by a NV center with frequency $\Omega_1$. In (a)-(d) the static in-plane magnetic field is rotated with $\theta=\{0, \pi, \pi/4, \pi/2\}$, respectively. The arrows represent the directions of $\mathbf{S}_{\bf q}$, and the color bar denotes their $|\mathbf{S}_{\bf q}|$. Parameters are given in the text.}
\label{spin_NV}
\end{figure}

Figure~\ref{spin_NV} plots the spin density $\mathbf{S}^{(n=1)}_{\bf q}$ for the in-plane-wave components of the stray magnetic field for a static magnetic field with different in-plane directions.
The stray magnetic field is always in-plane and perpendicular to the wave vector. The anticlockwise rotation of the stray field in the wave-vector space implies that the chirality of the dipolar field is always \textquotedblleft
right-handed\textquotedblright; in other words, the photon spin is locked to its wave vector. 

\section{Nonreciprocal quantum entanglement of NV spin qubits}

\subsection{Chiral coupling of magnons with NV spin qubit}

\label{magnetic_nanostructure}

\begin{figure}[htp!]
\centering
\includegraphics[width=0.45\textwidth,trim=0.6cm 0cm 0cm 0.1cm]{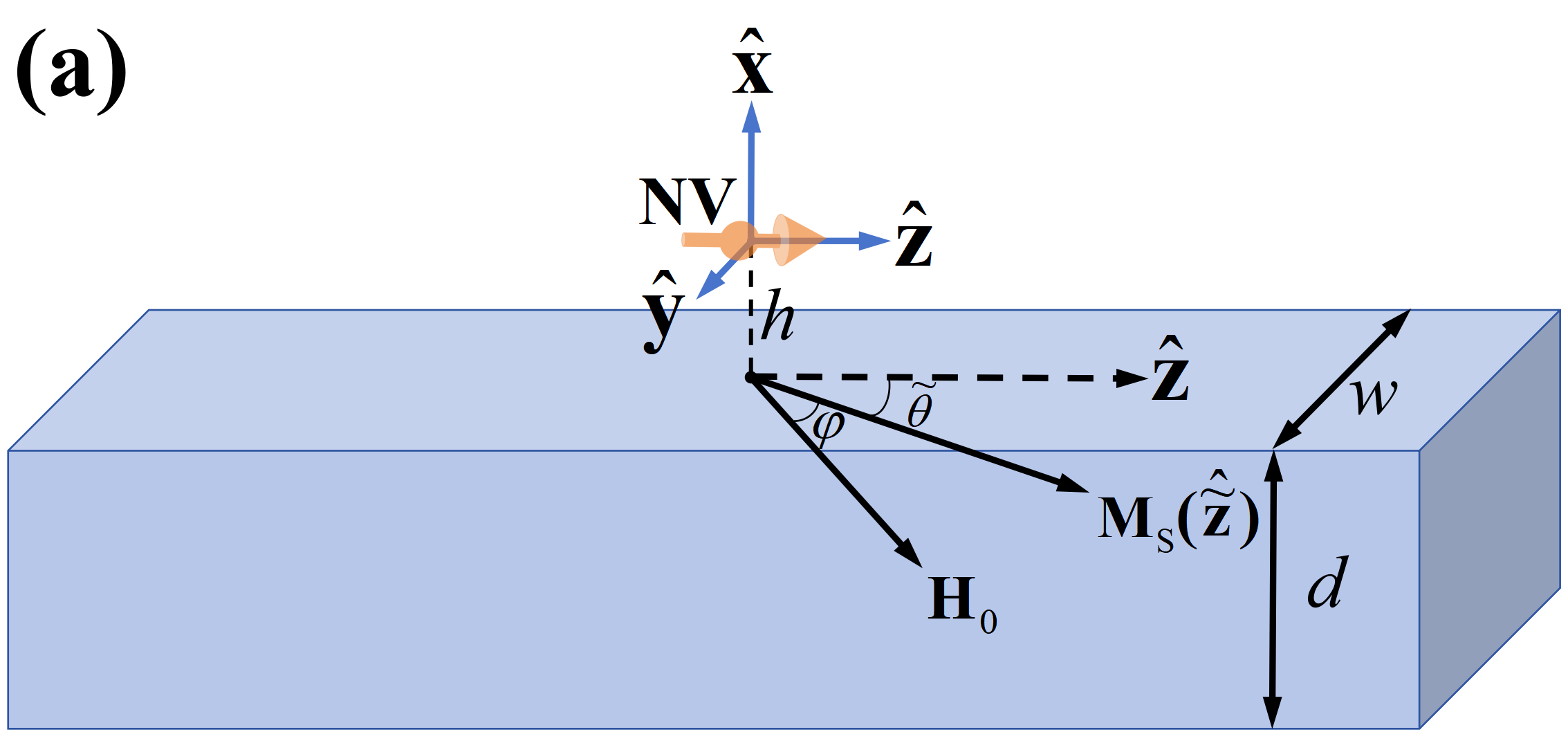}
\includegraphics[width=0.45\textwidth,trim=0.6cm 0cm 0cm 0.1cm]{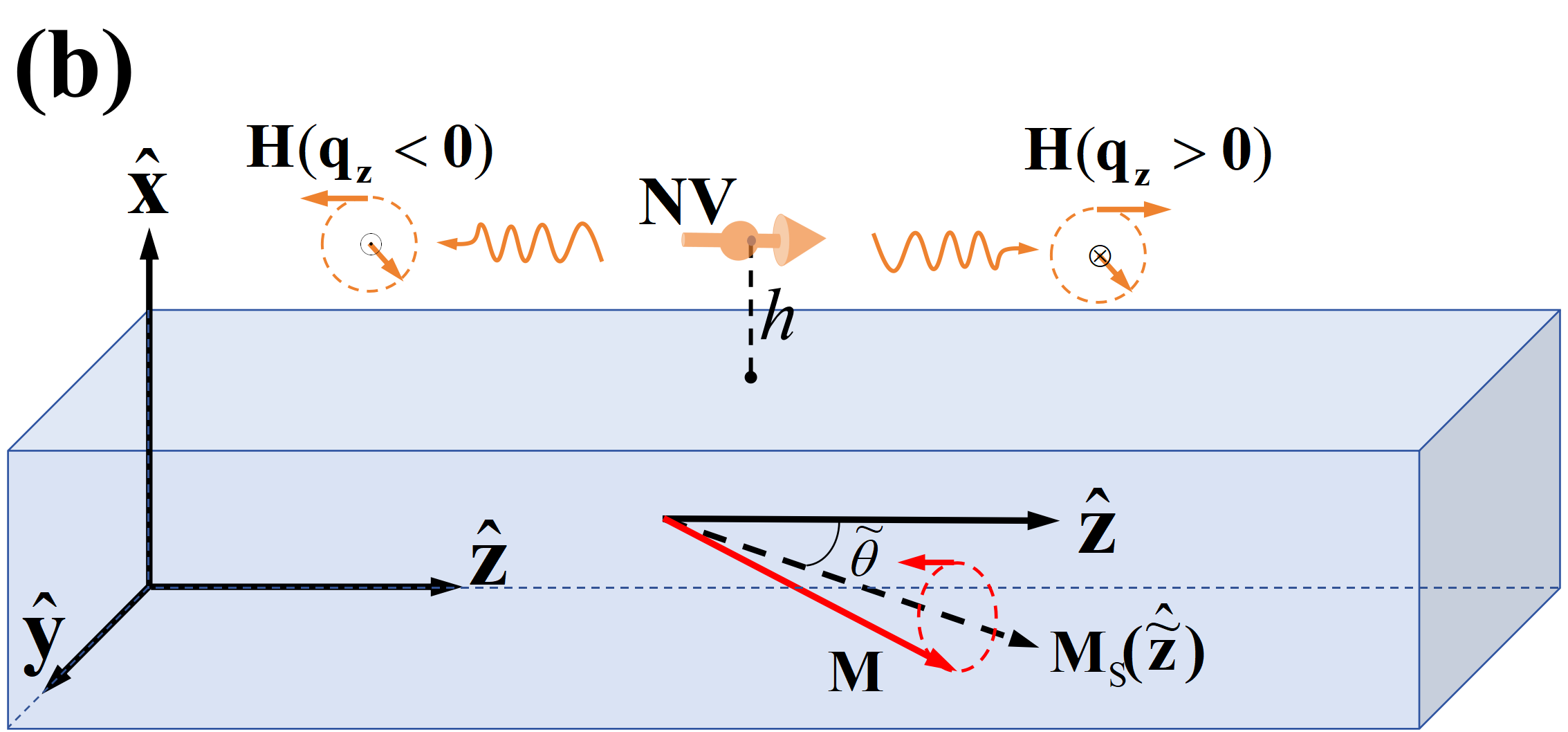}
\caption{(a): An NV center in diamonds at a vertical distance $h$ above the central axis of a magnetic wire along the $\hat{\bf z}$-direction. The origin in the text is shifted to the NV center. (b) illustrates the polarization-momentum locking of the stray field generated by an excited NV center with Fourier components $|q_y|\ll |q_z|$.}
\label{wire}
\end{figure}

Due to the quantum nature of the NV spin excitation, the transfer of the spin of the NV center by photons may excite one to a few magnons in a magnetic nanostructure beneath the NV centers at a small distance $h$. Here, we explore the unidirectional excitation of magnons in a magnetic nanowire of thickness $d$ and width $w$ along the $\hat{\bf z}$-direction. We allow the saturation magnetization ${\bf M}_s$ to follow the in-plane static magnetic field ${\bf H}_0$ that is set at an angle $\theta$ with respect to the wire $\hat{\bf z}$-direction.
As illustrated in Fig.~\ref{wire}(a), the saturation magnetization ${\bf M}_s$ has the direction $\tilde{\theta}=\theta-\varphi$, where $\varphi$ is the angle between ${\bf H}_0$ and ${\bf M}_s$, which minimizes the free-energy density
\begin{align}
    F_m=-\mu_0M_sH_0\cos(\theta-\tilde{\theta})+\frac{\mu_0}{2}N_{yy}M_s^2\sin^2\tilde{\theta},
\end{align}
where $N_{yy}\simeq d/(d+w)$ is the demagnetization factor of the wire~\cite{chiral_excitation}.
At the minimum,  $dF_m/d{\tilde{\theta}}=0$ and $d^2F_m/d{\tilde{\theta}}^2>0$, such that 
\begin{align}
    &\mu_0M_sH_0\sin(\theta-\tilde{\theta})-\mu_0 N_{yy}M_s^2\sin\tilde{\theta}\cos\tilde{\theta}=0,\nonumber\\
    &\mu_0M_sH_0\cos(\theta-\tilde{\theta})+\mu_0 N_{yy}M_s^2(\cos^2\tilde{\theta}-\sin^2\tilde{\theta})>0,
    \label{directions}
\end{align}
from which $\tilde{\theta}$ can be obtained numerically. 

We solve the eigenmodes ${\cal M}_{\beta}(k_z)$ of the magnetic order under the equilibrium configuration in Fig.~\ref{wire}(a) in Appendix~\ref{eigenmodes_nanowire}. In the second quantization, the magnetization operator with Cartesian components $\beta\in\{x,y,z\}$  reads
\begin{align}
    \hat{M}_{\beta}({\bf r})&=-\sqrt{2M_s\gamma\hbar}\sum_{k_z}{\cal M}_{\beta}(k_z) e^{ik_zz-i\omega_{k_z}t}\hat{\beta}_{k_z}+{\rm H.c.},
    \label{magnetization_operator}
\end{align}
where $\hat{\beta}^{\dagger}_{k_z}$ and $\hat{\beta}_{k_z}$ are the magnon creation and annihilation operators.

The magnetization couples with the stray field emanating from the NV center by the Zeeman interaction 
\begin{align}
    \hat{H}_{\rm int}=-\mu_{0}\int \hat{M}_{\beta}({\bf r})\hat{H}_{\beta}({\bf r})d{\bf r}.
    \label{interaction_Hamiltonian}
\end{align}
Substituting Eqs.~\eqref{stray_field_NV_operator} and  \eqref{magnetization_operator} and disregarding the rapidly oscillating terms yields  
\begin{align}
    \hat{H}_{\rm int}=\sum_{k_z}\left(\hbar g_{k_z}\hat{\sigma}^+\hat{\beta}_{k_z}+{\rm H.c.}\right),
\end{align}
with a coupling constant
\begin{align}
g_{k_z}
&=\frac{\mu_{0}\gamma\sqrt{2M_{s}\gamma\hbar}}{\hbar}\left(e^{-h|k_z|}-e^{-(d+h)|k_z|}\right)\nonumber\\
&\times\left(\mathcal{M}_z^{k_z}{S}_{eg}^z-\mathcal{M}_x^{k_z}{S}_{eg}^x+\frac{ik_z}{|k_z|}\left(\mathcal{M}_{x}^{k_z}{S}_{eg}^z+\mathcal{M}_{z}^{k_z}{S}_{eg}^x\right)\right).
\label{coupling_constant}
\end{align}
For opposite \(k_z\), 
\begin{subequations}
\begin{align}
    g_{k_z>0}&=-\frac{\mu_{0}\gamma\sqrt{2M_{s}\gamma\hbar}}{\hbar}\left(e^{-h|k_z|}-e^{-(d+h)|k_z|}\right)\nonumber\\
    &\times\mathcal{M}_{x}^{k_z}\left(1-\frac{\sin{\tilde{\theta}}}{{\cal D}(k_z)}\right)\left(-i{S}_{eg}^{z}+S_{eg}^{x}\right),\\
    g_{k_z<0}&=-\frac{\mu_{0}\gamma\sqrt{2M_{s}\gamma\hbar}}{\hbar}\left(e^{-h|k_z|}-e^{-(d+h)|k_z|}\right)\nonumber\\
&\times\mathcal{M}_{x}^{k_z}\left(1+\frac{\sin{\tilde{\theta}}}{{\cal D}(k_z)}\right)\left(i{S}_{eg}^z+S_{eg}^x\right).
\end{align}
\end{subequations}
The directionality 
\begin{align}
    \frac{|g_{k_z>0}|}{|g_{k_z<0}|}&=\frac{|({{\cal D}(k_z)}-{\sin{\tilde{\theta}}})(S_{eg}^z+iS_{eg}^x)|}{|({{\cal D}(k_z)}+{\sin{\tilde{\theta}}})(S_{eg}^z-iS_{eg}^x)|}\nonumber\\
    &=\frac{|({\cal M}_{z}^{k_z}+i{\cal M}_{x}^{k_z})(S_{eg}^z+iS_{eg}^x)|}{|({\cal M}_{z}^{k_z}-i{\cal M}_{x}^{k_z})(S_{eg}^z-iS_{eg}^x)|}
\end{align}
is large when $\tilde{\theta}$ is around $\pi/2$ and vanishes when $\tilde{\theta}=0$. 
The spin density of the stray magnetic field emitted by the NV center is along the negative (positive) $\hat{\bf y}$-direction when $k_z>0$ ($k_z<0$), as shown in Fig.~\ref{spin_NV}, i.e., the spin $S_y$ is locked to the wave vector $k_z$. 
On the other hand, the magnetization in the laboratory system 
\begin{align}
    {\cal M}(t)&=\left(\begin{array}{c}
         {\cal M}_{x}e^{-i\omega_{k_z}t}  \\
         {\cal M}_{y}e^{-i\omega_{k_z}t}     \\
         {\cal M}_{z}e^{-i\omega_{k_z}t}
    \end{array}\right)\propto e^{-i\omega_{k_z}t}\left(\begin{array}{c}
         \sqrt{\frac{{\cal D}(k_z)}{4dw}}  \\
         i\cos{\tilde{\theta}}\sqrt{\frac{1}{4dw{\cal D}(k_z)}}     \\
         -i\sin{\tilde{\theta}}\sqrt{\frac{1}{4dw{\cal D}(k_z)}}
    \end{array}\right)\nonumber\\&=\left(\begin{array}{ccc}
             1 & 0 & 0\\
            0 & \cos{\tilde{\theta}}&\sin{\tilde{\theta}} \\
             0 & -\sin{\tilde{\theta}} & \cos{\tilde{\theta}}
         \end{array}\right)\left(
         \begin{array}{c}
            \sqrt{\frac{{\cal D}(k_z)}{4dw}}   \\
               i\sqrt{\frac{1}{4dw{\cal D}(k_z)}}
              \\ 0
         \end{array}
         \right)e^{-i\omega_{k_z}t}\nonumber\\&\propto\left(\begin{array}{ccc}
             1 & 0 & 0\\
            0 & \cos{\tilde{\theta}}&\sin{\tilde{\theta}} \\
             0 & -\sin{\tilde{\theta}} & \cos{\tilde{\theta}}
         \end{array}\right)\tilde{\cal M}(t)
\end{align} 
with the real part
\begin{align}
    &{\rm Re}[\tilde{\cal M}(t)]\nonumber\\
    &\propto\left(\sqrt{\frac{{\cal D}(k_z)}{4dw}} \cos(\omega_{k_z}t) ,\sqrt{\frac{1}{4dw{\cal D}(k_z)}}\sin(\omega_{k_z}t),0\right)^T,
\end{align}
which precesses elliptically around $\tilde {\theta}$ as illustrated in Fig.~\ref{wire}(b). When $\tilde{\bf z}$ is along $\hat{\bf y}$, the circular polarization of the stray fields of the NV center propagating along the negative $\hat{\bf z}$-direction matches the spin waves propagating in the same direction, which explains the chiral coupling.

One excited NV spin qubit can excite propagating magnons by the chiral coupling as addressed in Appendix~\ref{appendix}. A second NV spin qubit separated from the first one can be excited by the propagating magnons, which mediate an effective interaction between the two NV spin qubits.

\subsection{Quantum entanglement}

\label{nonreciprocal_entanglement}

The  Hamiltonian $\hat{H}=\hat{H}_{\rm{NV}}+\hat{H}_{\rm {m}}+\hat{H}_{\rm{int}}$ of two identical NV centers $i=\{1,2\}$, located at $(0,0,z_i)$ on top of the magnetic nanowire, lives in the 4-dimensional subspace spanned by $\{|i,e\rangle,|i,g\rangle\}$:
\begin{subequations}
\begin{align}
    \hat{H}_{\rm{NV}}&=\sum_{i=\{1,2\}}\left(\hbar \omega_{e}|i,e\rangle \langle i,e|+\hbar \omega_{g}|i,g\rangle \langle i,g|\right),\\
    \hat{H}_{\rm {m}}&=\sum_{k_z}\hbar\tilde{\omega}_{k_z}\hat{\beta}_{k_z}^{\dagger}\hat{\beta}_{k_z}, 
\end{align}
\end{subequations}
where we augment the frequency $\tilde{\omega}_{k_z}=\omega_{k_z}(1-i\alpha_G)$ by the Gilbert damping $\alpha_G$.
In the rotating-wave approximation, the interaction reads 
\begin{align}
\hat{H}_{\rm{int}} 
\approx\sum_{i=\{1,2\}}\sum_{k_z}
\hbar g_{k_z}e^{ik_z z_i}\hat{\sigma}_{i}^{+}\hat{\beta}_{k_z}+\mathrm{H.c.},
\label{interaction_Hamiltonian}
\end{align}
where $\hat{\sigma}_{i}^{+}=|i,e\rangle \langle i,g|$, $\hat{\sigma}_{i}^{-}=|i,g\rangle \langle i,e|$, and the coupling constant $g_{k_z}$ is given by Eq.~\eqref{coupling_constant}. The direct interaction between the NV centers may be disregarded when separated by more than 30~nm \cite{Dolde}.

The master equation for the density matrix $\hat{\rho}^I$ of the NV centers \textcolor{blue}{defined in the interaction picture with the interaction Hamiltonian given by Eq.~\eqref{interaction_Hamiltonian}} reads \textcolor{blue}{(refer to Appendix~\ref{derivation_master_equation} for the derivation of the master equation with  justification of the Born-Markov
approximation)}
\begin{align}
    \frac{d \hat{\rho}^{I}}{d t}&=-i\left[\sum_{\{i,j\}=1,2}\Omega_{ij}\hat{\sigma}_{j}^{+}\hat{\sigma}_{i}^{-},\hat{\rho}^{I}\right]\\
    &+i\left[\Omega\sum_{i={1,2}}\hat{\sigma}^{z}_{i},\hat{\rho}^{I}\right]
    -\Omega_{\alpha}\left[\sum_{i=1,2}\hat{\sigma}^z_{i},\hat{\rho}^I\right]\nonumber\\
    &-\sum_{i,j}\Omega^{\downarrow}_{ ij}\left(\hat{\sigma}_{j}^{+}\hat{\sigma}_{i}^{-}\hat{\rho}^{I}+\hat{\rho}^{I}\hat{\sigma}_{j}^{+}\hat{\sigma}_{i}^{-}-2\hat{\sigma}_{i}^{-}\hat{\rho}^{I}\hat{\sigma}_{j}^{+}\right)\nonumber\\
    &-\sum_{i,j}\Omega^{\uparrow}_{ij}\left(\hat{\sigma}_{i}^{-}\hat{\sigma}_{j}^{+}\hat{\rho}^{I}+\hat{\rho}^{I}\hat{\sigma}_{i}^{-}\hat{\sigma}_{j}^{+}-2\hat{\sigma}_{j}^{+}\hat{\rho}^{I}\hat{\sigma}_{i}^{-}\right),\nonumber\\
    &\approx-i\left(\hat{H}_{\rm eff}\hat{\rho}^{I}-\hat{\rho}^{I}\hat{H}_{\rm eff}^{\dagger}\right)-\Omega_{\alpha}\left[\sum_{i=1,2}\hat{\sigma}^z_{i},\hat{\rho}^I\right]\nonumber\\
    &+\sum_{i,j}2\Omega^{\downarrow}_{ ij}\hat{\sigma}_{i}^{-}\hat{\rho}^{I}\hat{\sigma}_{j}^{+}+\sum_{i,j}2\Omega^{\uparrow}_{ ij}\hat{\sigma}_{j}^{+}\hat{\rho}^{I}\hat{\sigma}_{i}^{-},
    \label{master_equation}
\end{align}
with coefficients
\begin{align}
\Omega_{ij}&=
     \sum_{k_z}|g_{k_z}|^{2}e^{i k_z(z_{j}-z_{i})}{\rm Re}\left(\frac{1}{\omega_{\rm NV}-\tilde{\omega}_{k_z}}\right),\nonumber\\
     \Omega&=-\sum_{k_z}|g_{k_z}|^{2}{\rm Re}\left(\frac{1}{\omega_{\rm NV}-\tilde{\omega}_{k_z}}\right)n(\hbar\omega_{\rm NV}),\nonumber\\
    \Omega^{\downarrow}_{ij}&=\gamma_{ij}(\omega_{\rm NV})\left(n(\hbar\omega_{\rm NV})+1\right),\nonumber\\
    \Omega^{\uparrow}_{ij}&=\gamma_{ij}(\omega_{\rm NV})n(\hbar\omega_{\rm NV}),\nonumber\\
    \Omega_{\alpha}&={2\pi}/{\tau},
    \label{coefficients}
\end{align}
where $\gamma_{ij}(\omega_{\rm NV})=\sum_{ k_z}\pi|g_{k_z}|^{2}\delta(\omega_{k_z}-\omega_{\rm NV})e^{ik_z(z_{j}-z_{i})}$, $n(\hbar \omega_{\rm NV})=1/\{\exp[\hbar \omega_{\rm NV}/(k_BT)]-1\}$ is the Bose-Einstein distribution at temperature $T$, and $\tau$ is the dephasing time of the NV centers.   $\Omega_{ii}^\downarrow$ denotes the relaxation rate induced by the magnetic nanowire, which is enhanced by the temperature, while  $\Omega_{ii}^\uparrow$ describes the reverse process with $\Omega_{ii}^\uparrow= e^{-\beta\hbar \omega_{\text{NV}}}\Omega_{ii}^\downarrow$. The correlations in the magnetic wire also give rise to the collective decay $\Omega_{12}^\downarrow$ and its reverse process $\Omega_{12}^\uparrow$~\cite{yu2024non,zou2022bell,zou2024spatially}.

From the quantum master equation, we obtain an effective Hamiltonian for the two NV centers
\begin{align}
    \hat{H}_{\rm eff}&=\sum_{i,j=1,2}\left[(\Omega_{ij}-{i}\Omega_{ij}^{\downarrow}-{i}\Omega_{ij}^{\uparrow})\hat{\sigma}_{j}^{+}\hat{\sigma}_{i}^{-}+{i}\Omega_{ij}^{\uparrow}\delta_{ij}\hat{\sigma}_{i}^{z}-\Omega\hat{\sigma}^{z}_{i}\right]\nonumber\\
    &\overset{T\rightarrow 0}{=}\sum_{i,j=1,2}\Gamma_{ij}\hat{\sigma}_{j}^{+}\hat{\sigma}_{i}^{-},
    \label{effective_Hamiltonian}
\end{align}
in which at zero temperature, we obtain an effective coupling between the NV centers
\begin{align}
    \Gamma_{ij}=
     \sum_{k_z}|g_{k_z}|^{2}\frac{e^{i k_z(z_{j}-z_{i})}}{\omega_{\rm NV}-\tilde{\omega}_{k_z}}.
\end{align}
The local term $\Gamma_{ii}$ represents a Lamb-like frequency shift of the NV centers, induced by virtual magnon fluctuations. Physically, this shift arises from the dressing of the NV energy levels by the vacuum fluctuations of the magnon field, even in the absence of real magnon excitations. The correlations in the magnetic wire also couple the \textcolor{blue}{NV centers} by  $\Gamma_{12}$ and $\Gamma_{21}$.
From the effective Hamiltonian \eqref{effective_Hamiltonian}, for a single NV center $\hat{H}_{\rm eff}=(\Omega_{11}-{i}\Omega_{11}^{\downarrow}-{i}\Omega_{11}^{\uparrow})\hat{\sigma}_{1}^{+}\hat{\sigma}_{1}^{-}+{i}\Omega_{11}^{\uparrow}\hat{\sigma}_{1}^{z}$, in which the temperature enhances its damping/fluctuation and shifts the frequency of the states.

\begin{figure}[htp!]
\centering\includegraphics[width=0.85\linewidth]{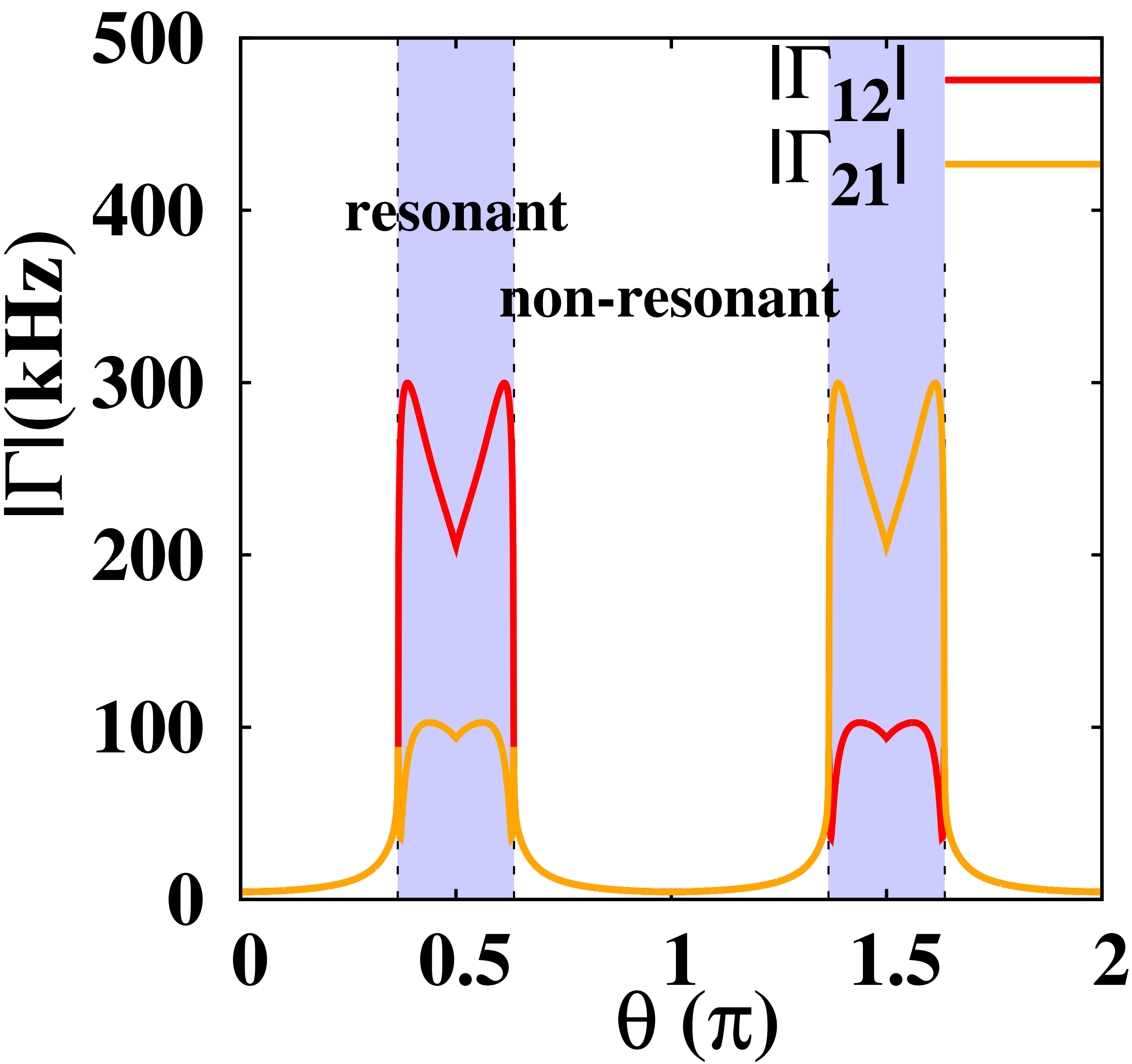}
\caption{Coupling constants $\Gamma_{12}$ and $\Gamma_{21}$ between two NV-centers above a magnetic nanowire as a function of the applied magnetic field angle $\theta$ (details are given in the text).
\label{Gamma}
}
\label{NV_coupling}
\end{figure}

We assume the diameter of the diamond nanoparticle is small such that there is one NV center per diamond nanoparticle~\cite{CoherentGQ,GaebelNP,MazeNature,Hong,Maletinsky,CooperPRL,Thiel,Gieseler,Conangla,Schirhagl,FincoNC,Huxter} or a diamond film with a low density of NV centers in which a single pair interacts with the magnetic wire. The excitation efficiency is high when the equilibrium magnetization lies parallel to the wire direction. 
A sufficiently weak applied magnetic field such as $\mu_0H_0=0.05$~T then cannot tilt the magnetization significantly, so  \(-20^{\circ}<\tilde{\theta} <20^{\circ} \) for all \(\theta\). 
When adopting a thickness $d=35$~nm, width $w=10$~nm, low temperature saturation magnetization $\mu_0M_s=0.21$~T, Gilbert damping $\alpha_G=10^{-4}$, and
exchange stiffness $\alpha_{\mathrm{ex}}=3\times
10^{-16}~\mathrm{m}^{2}$, the frequency of the NV center $\Omega_{\rm NV}$ lies in the magnon band continuum when the in-plane magnetic field is applied at an angle $\theta$ around $\pi/2$ and $3\pi/2$, as shown in Fig.~\ref{excitation1d}(a) of Appendix~\ref{appendix}. 
The distance between the two NV centers is $|z_1-z_2|=100$~nm, separated from the nanowire surface by $h=15$~nm. \textcolor{blue}{The dephasing time of the individual NV center $\tau=1$~ms~\cite{Herbschleb}. We  solve the master equation for a low temperature of $T=100$~mK without driving field.}
Figure~\ref{NV_coupling} plots the calculated coupling constants $\Gamma_{12}$ and $\Gamma_{21}$ as a function of the magnetic field angle $\theta$.
In the gray region the interaction is ``resonant" with $\omega_{\rm NV}>\omega_{k_z}$, while it is ``non-resonant"  with $\omega_{\rm NV}<\omega_{k_z}$, otherwise. 
We see that in the non-resonant regime $|\Gamma_{12}|=|\Gamma_{21}|$ is relatively weak.  The coupling in the resonant regime is much larger, as is the difference between $\Gamma_{12}$ and $\Gamma_{21}$, which we refer to as a chiral coupling.

We now address the quantum entanglement between the coupled color centers with emphasis on the chiral coupling that causes nonreciprocal quantum entanglement between two NV spin qubits, which strongly depends on which NV center is initially excited.  We propose that it acts as an isolator in the formation of the quantum-entanglement~\cite{chiral_quantum_optics}, which is an analogue with an optical isolator that suppresses unwanted stray fields in optical networks.

We characterize the time-dependent quantum entanglement of the two-qubit state, obtained from the solution of the quantum master equation~\eqref{master_equation}, using the concurrence $C$~\cite{concurrence_1}.
For a general (mixed) state $\rho$, the concurrence is defined via the spectrum of $\rho \tilde{\rho}$, where $\tilde{\rho} \equiv (\sigma_y \otimes \sigma_y) \rho^* (\sigma_y \otimes \sigma_y)$ is the spin-flipped (or time-reversed) state, and $\rho^*$ denotes the complex conjugation in the computational basis. Specifically,
\begin{align}
C(\rho) = \max\{0, \lambda_1 - \lambda_2 - \lambda_3 - \lambda_4\},
\label{definition_concurrence}
\end{align}
where $\lambda_i$ are the square roots of the eigenvalues of $\rho \tilde{\rho}$ in the descending order. 
This construction quantifies the degree to which $\rho$ coherently overlaps with its spin-flipped counterpart. The concurrence takes values from 0 (for separable states) to 1 (for maximally entangled states), providing a normalized measure of two-qubit entanglement. 
In our case, for the two NV centers, the density matrix
\begin{align}
    \hat{\rho}=(|e,e\rangle,|e,g\rangle,|g,e\rangle,|g,g\rangle)\left(\begin{array}{cccc}
      \rho_{11} & \rho_{12} & \rho_{13} & \rho_{14} \\
        \rho_{21} & \rho_{22} & \rho_{23} & \rho_{24} \\ 
       \rho_{31} & \rho_{32} & \rho_{33} & \rho_{34} \\
     \rho_{41} & \rho_{42} & \rho_{43} & \rho_{44}  \\
    \end{array}\right)\left(\begin{array}{c}
         \langle e,e|  \\
         \langle e,g| \\
         \langle g,e| \\
         \langle g,g| 
    \end{array} \right).
\end{align}
With the initial state of the density matrix is $\hat{\rho}(t=0)=|e,g\rangle\langle e,g|$ or $\hat{\rho}(t=0)=|g,e\rangle\langle g,e|$, the density matrix at time $t$ is in the form of 
\begin{align}
    {\rho}(t)&=\left(\begin{array}{cccc}
        0 & 0 &0 &0\\
        0 & \rho_{22} & \rho_{23} & 0\\ 0 &\rho_{32} &\rho_{33} & 0\\ 0 & 0 & 0 &\rho_{44}
    \end{array}\right),
\end{align}
which by Eq.~\eqref{definition_concurrence} leads to the concurrence 
\begin{align}
    C(\rho(t))=2|\rho_{23}(t)|.
\end{align}

\begin{figure}
    \centering
    \includegraphics[width=0.75\linewidth]{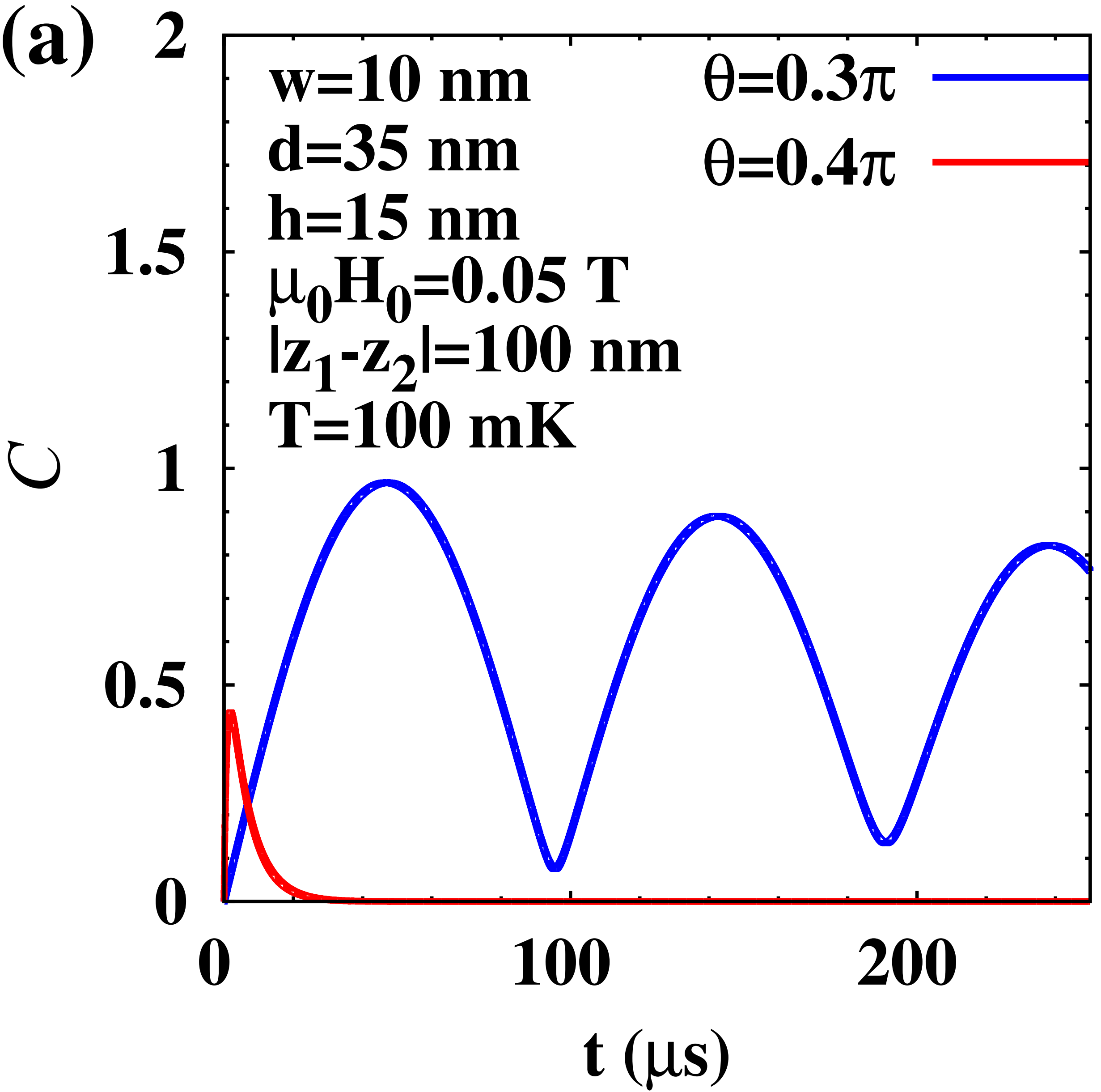}
    \hspace{-0.23cm}
    \includegraphics[width=0.755\linewidth]{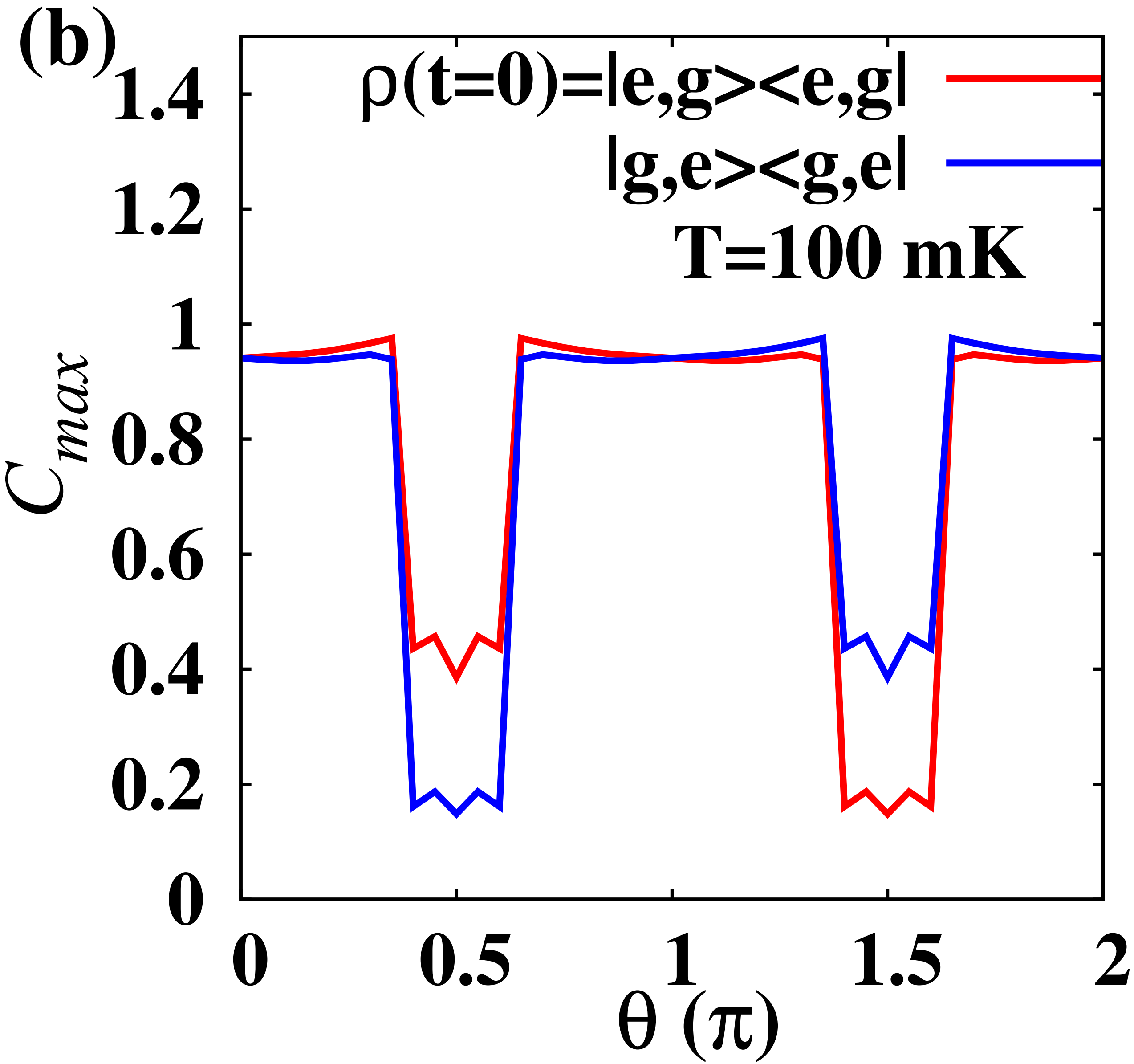}\hspace{-0.2cm}
    \caption{Quantum-entanglement isolator. Two NV spin qubits with a horizontal distance $|z_1-z_2|=100$~nm are coupled through magnons in a YIG nanowire (thickness $d=35$~nm, width $w=10$~nm), both of which are located $h=15$~nm above the nanowire. \textbf{(a)}: Evolution of the concurrence $C$ of two coupled NV spin qubits in the non-resonant and resonant regimes with initial state $\hat{\rho}(t=0)=|e,g\rangle\langle e,g|$.\textbf{ (b)}: Comparison of the maximal concurrence $C_{\rm max}$ in the non-resonant and resonant regimes with different initial states $\hat{\rho}(t=0)=|e,g\rangle\langle e,g|$ and $\hat{\rho}(t=0)=|g,e\rangle\langle g,e|$.}
    \label{entanglement}
\end{figure}

Figure~\ref{entanglement}(a) is a plot of the calculated concurrence when the left NV spin qubit is initially in its excited state with $\hat{\rho}(t=0)=|e,g\rangle\langle e,g|$ and parameters as in Fig.~\ref{Gamma}. 
The concurrence/quantum entanglement of the non-resonant and resonant regimes is very different. In the non-resonant regime, the effective coupling between the two NV centers is small (Fig.~\ref{Gamma}) and the quantum entanglement decays slowly with Rabi oscillation, but takes a much larger formation time. By contrast, in the resonant regime, the effective coupling is nonreciprocal $|\Gamma_{12}|\ne |\Gamma_{21}|$ and much enhanced compared to the non-resonant regime. In this case, the formation of the quantum entanglement is much faster (within a microsecond) and decays without oscillation, with a lifetime achieving the order of microseconds, and the maximal concurrence is not suppressed much compared to the coherent resonant regime.

Moreover, it turns out that the formed quantum entanglement in the resonant regime strongly depends on which NV spin qubit is initially in its excited state, as addressed in Fig.~\ref{entanglement}(b). The formed quantum entanglement differs significantly when  $\hat{\rho}(t=0)=|e,g\rangle\langle e,g|$ (the left NV spin qubit is initially excited) and $\hat{\rho}(t=0)=|g,e\rangle\langle g,e|$ (the right NV spin qubit is initially excited), which acts as excellent isolator for the formation of quantum entanglement.

\section{Conclusion and discussion}

\label{summary}

In conclusion, we attribute the chiral coupling between the excitation of the NV spin qubit and the spin waves in ferromagnets to the photonic spin-orbit coupling of the near stray field of the NV-center excitations, which we show to be always perfectly right-handed with chirality index $(-\hat{\bf x})\cdot(\hat{\bf S}_{\bf q}\times\hat{\bf q})=1$. An excited NV spin qubit can therefore serve as a source of propagating magnons with fixed frequency and wave vector. A propagating wave packet excited by the stray field of an NV center in diamond may generate a tunable nonreciprocal quantum entanglement when interacting with a distant second NV center,  with lifetimes only governed by the Gilbert damping. The nonreciprocity of quantum entanglement has the functionality of an isolator, since it may suppress unwanted communication between qubits by stray entanglement in a quantum network.

\begin{acknowledgments}
This work is financially supported by the National Key Research and Development Program of China under Grant No.~2023YFA1406600 and the National Natural Science Foundation of China under Grant No.~12374109. J.Z. acknowledges the support of the Georg H. Endress Foundation. The JSPS KAKENHI Grants No.~19H00645, 22H04965, and JP24H02231 support G.B. financially.
\end{acknowledgments}

\begin{appendix}

\section{Evolution of NV-spin driven by microwaves}
\label{photon_spin}

A wave function $|\psi(t)\rangle=C_g(t)|g\rangle+C_e(t)|e\rangle+C_f(t)|f\rangle$ of a NV center evolves according to 
\begin{align}
    i\left(\begin{matrix}
    \dot{C}_f \\
    \dot{C}_e \\
    \dot{C}_g
    \end{matrix}\right)=\sin(\omega_{{\rm ext}}t){\cal G}\left(\begin{matrix}
     C_f \\
    C_e \\
    C_g
    \end{matrix}\right)+\left(\begin{matrix}
      \omega_f & 0  & 0   \\
      0& \omega_e & 0   \\
      0 & 0 & \omega_g 
\end{matrix}\right)\left(\begin{matrix}
    C_f \\
    C_e \\
    C_g
    \end{matrix}\right).
    \label{coefficient_equation}
\end{align}
In the interaction picture the amplitudes $C_f(t)=C^I_f(t)e^{-i\omega_f t}$,  $C_e(t)=C^I_e(t)e^{-i\omega_e t}$, and $C_g(t)=C^I_g(t)e^{-i\omega_g t}$ obey
\begin{align}
    &i\left(\begin{matrix}
     \dot{C}^I_f(t) \\
    \dot{C}^I_e(t) \\
    \dot{C}^I_g(t)
\end{matrix}\right)=-\frac{i}{2}\left(e^{i\omega_{{\rm ext}}t}-e^{-i\omega_{{\rm ext}}t}\right){\cal G}^I\left(\begin{matrix}
     C^I_f(t) \\
    C^I_e(t) \\
    C^I_g(t)
\end{matrix}\right),
\label{coefficient_equation1}
\end{align}
where ${\cal G}^I_{ij}={\cal G}_{ij}e^{i(\omega_i-\omega_j)t}$.

We assume nearly resonant excitation with photon frequency $\omega_{{\rm ext}}\simeq\omega_{{\rm NV}}=\omega_e-\omega_g$ and adopt the rotating-wave approximation. To leading order in the small parameter $\mu_0\gamma h_0\ll \omega_{{\rm ext}}$, Eq.~(\ref{coefficient_equation1}) simplifies to
\begin{align}
     i\frac{d}{dt}\left(\begin{matrix}
    C^I_e(t) \\
    C^I_g(t)
\end{matrix}\right)=\left(\begin{matrix}
     0 & \frac{i}{2}G_{eg}   \\
    -\frac{i}{2} G_{ge} & 0 
\end{matrix}\right)\left(\begin{matrix}
    C^I_e(t) \\
    C^I_g(t)
\end{matrix}\right).
\label{coefficient_equation2}
\end{align}
The general solution of the coupled equation
\begin{align}
    \ddot{C}^I_g(t)+|G_{eg}|^2 C^I_g(t)/4=0,\nonumber\\
    \ddot{C}^I_e(t)+|G_{eg}|^2 C^I_e(t)/4=0,
    \label{coupled_equations}
\end{align}
reads
\begin{align}
    C_{g,\eta}(t)&=A_{\eta}e^{-i\eta\frac{|G_{eg}|}{2}t}e^{-i\omega_g t},\nonumber\\
    C_{e,\eta}(t)&=B_{\eta}e^{-i\eta\frac{{|G_{eg}|}}{2}t}e^{-i\omega_e t},
\end{align}
\textcolor{blue}{where $\eta=\pm$  and the coefficients $\{A_{\eta},B_{\eta}\}$ follow from $ A_{\eta}=i\eta (S_{ge}^{y*}/|S_{ge}^y|)B_{\eta}$ and the normalized condition $|A_{\eta}|^2+|B_{\eta}|^2=1$}.
The ground state, with $C_g(t=0)=1$ and $C_e(t=0)=0$,  evolves in time according to
\begin{subequations}
    \begin{align}
    C_g(t)=\cos{(|G_{eg}|t/2)}e^{-i\omega_g t},\\
    C_e(t)=\sin{(|G_{eg}|t/2)}e^{-i\omega_e t}.
\end{align}
\end{subequations}

The expectation values of the spin operator \eqref{Ssigma} 
\begin{align}
    S^j_{\rm NV}(t)&=\langle\psi(t)|\hat{S}^j_{\rm NV}|\psi(t)\rangle\nonumber\\
    &=\frac{1}{2}(S^j_{gg}+S^j_{ee})+\frac{1}{4}(S^j_{gg}-S^j_{ee})(e^{-i\Omega_3 t}+\mathrm{H.c.})\nonumber\\
    &+\left[\frac{i}{4}S^j_{ge}\left(e^{-i\Omega_1 t}-e^{-i\Omega_2 t}\right)+\mathrm{H.c.}\right]
    \label{spin_components}
\end{align}
oscillate with three characteristic frequencies $\Omega_1=\omega_{\rm NV}+|G_{eg}|$, $\Omega_2=\omega_{\rm NV}-|G_{eg}|$, and $\Omega_3=|G_{eg}|$, as illustrated in Fig.~\ref{energy_level_splitting}, which can be interpreted in terms of the emission and absorption of a microwave photon by the NV center.
The modulus of the spin of the NV centers is also time-dependent.
When ${\bf H}_0\parallel \hat{\bf n}_{\rm NV}\parallel \hat{\bf z}$ and $D_{\rm NV}>\mu_0\gamma  H_0$, 
\begin{align}
    S_{\rm NV,\parallel}^{x}(t)&=\langle \psi|\hat{S}_{\rm NV}^x |\psi\rangle=\frac{\hbar}{\sqrt{2}}\sin(\frac{\mu_0\gamma h_0}{\sqrt{2}}t)\cos(\omega_{\rm NV}t)\nonumber\\
    &=\frac{i\hbar}{4\sqrt{2}}\left(e^{-i\Omega_1t}-e^{-i\Omega_2t}\right)+{\rm H.c.},  \nonumber\\
    S_{\rm NV,\parallel}^y(t)&=\langle \psi|\hat{S}_{\rm NV}^y |\psi\rangle=-\frac{\hbar}{\sqrt{2}}\sin(\frac{\mu_0\gamma h_0}{\sqrt{2}}t)\sin(\omega_{\rm NV}t)\nonumber\\
    &=\frac{\hbar}{4\sqrt{2}}\left(e^{-i\Omega_1t}-e^{-i\Omega_2t}\right)+{\rm H.c.}, \nonumber\\
    S_{\rm NV,\parallel}^z(t)&=\langle \psi|\hat{S}_{\rm NV}^z |\psi\rangle=-\frac{\hbar}{2}+\frac{\hbar}{2}\cos(\frac{\mu_0\gamma h_0}{\sqrt{2}}t)\nonumber\\
    &=-\frac{\hbar}{2}+\left(\frac{\hbar}{4}e^{-i\Omega_3t}+{\rm H.c.}\right).
    \label{spin_parallel}
    \end{align}
 When starting from the ground state with ${\bf S}_{\rm NV,\parallel}(t=0)=0$, the total spin $|{\bf S}_{{\rm NV},\parallel}(t)|=\hbar\sqrt{1-\cos^4({\mu_0\gamma h_0 t}/{2\sqrt{2}}})\le \hbar$ oscillates in time with a period that depends only on the frequency \(\mu_0\gamma h_0\). 

\section{Spin-wave eigenmodes of magnetic nanowires}
\label{eigenmodes_nanowire}

We solve the eigenmodes of the magnetic order under the equilibrium configuration in Fig.~\ref{wire}(a) 
in the local coordinate $\{\tilde{x},\tilde{y},\tilde{z}\}$ in which the $\tilde{\bf z}$-axis is along the saturation magnetization ${\bf M}_s$, starting from the Landau-Lifshitz-Gilbert (LLG) equation
\begin{align}
    \frac{\partial\tilde{\bf M}}{\partial t}=-\mu_0\gamma\tilde{\bf M}\times\tilde{\bf H}_{\rm eff}+\frac{\alpha_G}{M_s}\tilde{\bf M}\times\frac{\partial\tilde{\bf  M}}{\partial t},
    \label{LLG}
\end{align}
where the magnetization $\tilde{\bf M}=M_{\tilde{x}}\tilde{\bf x}+M_{\tilde{y}}\tilde{\bf y}+M_{\tilde{z}}\tilde{\bf z}$ and $\alpha_G$ is the Gilbert damping constant. The effective magnetic field $\tilde{\bf H}_{\rm eff}$ is the sum of the applied magnetic field $\tilde{\bf H}_0=H_0\sin{\varphi}\tilde{\bf y}+H_0\cos{\varphi}\tilde{\bf z}$, the exchange magnetic field $\tilde{\bf H}_{\rm ex}=\alpha_{\rm ex}\tilde{\nabla}^2\tilde{\bf M}$ with exchange stiffness $\alpha_{\rm ex}$, and the shape anisotropy field $\tilde{\bf H}_d=H_x^d\tilde{\bf x}+(H_y^d\cos\tilde{\theta}-H_z^d\sin\tilde{\theta})\tilde{\bf y}+(H_y^d\sin\tilde{\theta}+H_z^d\cos\tilde{\theta})\tilde{\bf z}$, where  $H_x^d=-N_{xx}M_x$, $H_y^d=-N_{yy}M_y$,  and $H_z^d=-N_{zz}M_z$ with the demagnetization factors $N_{xx}\simeq w/(d+w)$, $N_{yy}\simeq d/(d+w)$, and $N_{zz}\simeq 0$~\cite{chiral_excitation,efficient_gating}. 
After linearization in terms of small precession amplitudes and focusing on the lowest perpendicular standing wave mode for  \(\alpha_G=0\), the dispersion relation of the spin waves propagating along the wire direction with wave number $k_z$ 
\begin{align}
\omega_{k_z}=\sqrt{\omega_1(k_z)\omega_2(k_z)},
\end{align}
with
\begin{align}
&\omega_1=\mu_0\gamma\left(H_0\cos\varphi+N_{xx}M_s-N_{yy}M_s\sin^2\tilde{\theta}+\alpha_{\rm ex}k_z^2M_s\right),\notag\\
&\omega_2=\mu_0\gamma\left(H_0\cos\varphi+N_{yy}(\cos^2\tilde{\theta}-\sin^2\tilde{\theta})M_s+\alpha_{\rm ex}k_z^2M_s\right).
\end{align}
The spin-wave amplitudes read $M^{k_z}_{\tilde{x},\tilde{y}}({\bf r})={\cal M}_{\tilde{x},\tilde{y}}(k_z)G(x,y)$, where $G(x,y)=\Theta(-x-h)\Theta(x+d+h)\Theta(y+w/2)\Theta(-y+w/2)$ defines the wire cross section in terms of the Heavyside $\Theta(x)$ step function.
According to the linearized LL equation ${\cal M}_z^{k_z}=-i\sin\tilde{\theta}{\cal M}_x^{k_z}/{{\cal D}(k_z)}$, where
\begin{align}
    &{\cal D}(k_z)=\sqrt{{\omega_2}/{\omega_1}}\nonumber\\
    &=\sqrt{\frac{H_0\cos\varphi+N_{yy}(\cos^2\tilde{\theta}-\sin^2\tilde{\theta})M_s+\alpha_{\rm ex}M_sk_z^2}{H_0\cos\varphi+N_{xx}M_s-N_{yy}M_s\sin^2\tilde{\theta}+\alpha_{\rm ex}M_sk_z^2}}.
    \label{DD}
\end{align}
Implementing the normalization condition~\cite{magnetostatic_modes,spin_wave_excitation} $\int d{\bf r}(M^{k_z}_{\tilde{x}}({\bf r})M_{\tilde{y}}^{k_z*}({\bf r})-M^{k_z}_{\tilde{y}}({\bf r})M_{\tilde{x}}^{k_z*}({\bf r}))=-i/2$, we obtain the amplitudes normalized for a single magnon
\begin{align}
    {\cal M}_{\tilde{x}}(k_z)=\sqrt{\frac{{\cal D}(k_z)}{4wd}},~~~ {\cal M}_{\tilde{y}}(k_z)=i\sqrt{\frac{1}{4{\cal D}(k_z)wd}},
\end{align}
which in the lab coordinate system becomes
${\cal M}_x(k_z)={\cal M}_{\tilde{x}}(k_z)$, ${\cal M}_y(k_z)=\cos\tilde{\theta}{\cal M}_{\tilde{y}}(k_z)$, and ${\cal M}_z(k_z)=-\sin\tilde{\theta}{\cal M}_{\tilde{y}}(k_z)$.

\section{Excitation of magnons by excited NV spin qubits}

\label{appendix}

In this appendix, we address the quantum dynamics of magnons in the magnetic nanowire coupled to an excited NV center.

With the interaction Hamiltonian \eqref{interaction_Hamiltonian}, the equation of motion for the magnon operator $\hat{\beta}_{k_z}(t)$ in the frequency domain reads  
\begin{align}
    (\omega-\tilde{\omega}_{k_z})\langle\hat{\beta}_{k_z}(\omega)\rangle&=g_{k_z}^{*}\langle\hat{\sigma}^{-}(\omega)\rangle.
    \label{msigma}
\end{align} 
Under the resonant excitation with frequency $ \omega_{\rm NV}$, the expectation of the annihilation operator of the NV center for the wavefunction \eqref{general_solution}  
\begin{align}
     \sigma^{-}(t)&=\langle \psi(t)| \hat{\sigma}^{-}|\psi(t)\rangle=({i}/{4})\left(e^{-i\Omega_1 t}-e^{-i\Omega_2 t}\right),
     \nonumber\\
    \sigma^{-}(\omega)&=\int \sigma^{-}(t)e^{i\omega t}{dt}=\frac{i\pi}{2}\left(\delta(\omega-\Omega_1)-\delta(\omega-\Omega_2) \right),
    \label{anihilation_operator}
\end{align}
oscillates with the transition frequencies $\Omega_1$ and $\Omega_2$  in Fig.~\ref{energy_level_splitting}. 
$\Omega_3$ does not appear in the rotating-wave approximation.
Substituting \(\sigma^{-}(\omega)\) into Eq.~\eqref{msigma} leads to 
\begin{align}
    \beta_{k_z}(\omega)=\frac{g_{k_z}^{*}\langle\hat{\sigma}^{-}(\omega)\rangle}{\omega-\tilde{\omega}_{k_z}}=\frac{i\pi g_{k_z}^*\left(\delta(\omega-\Omega_1)-\delta(\omega-\Omega_2)\right)}{2(\omega-\tilde{\omega}_{k_z})}.
    \label{mkomega}
\end{align}
By transforming back into the time domain,  
\begin{align}
    \beta_{k_z}(t)&=\int\frac{d\omega}{2\pi} \beta_{k_z}(\omega)e^{-i\omega t}\nonumber\\
    &=\frac{ig_{k_z}^*}{4}\left(\frac{e^{-i\Omega_1t}}{\Omega_1-\tilde{\omega}_{k_z}}-\frac{e^{-i\Omega_2t}}{\Omega_2-\tilde{\omega}_{k_z}}\right).
    \label{excited_magnon}
\end{align}

In a diamond particle with tens of nanometers in diameter and a typical density of NV centers of $\sim 1\times 10^{11}~{\rm NV}/{\rm cm}^2$~\cite{Iacopo}, the resonant microwaves trigger  $N_{\rm NV} \ge 1$ NV centers with nearly the same $\hat{\bf n}_{\mathrm NV}$ and \(\omega_{\mathrm NV}\) simultaneously with stray-field amplitudes that interfere constructively. 
The number of magnons excited by one NV center reads
\begin{align}
    N(t)=\int \frac{dk_z}{2\pi} |\beta_{k_z}(t)|^2,  
    \label{Number}
    \end{align}
with a beating pattern due to the interference of the different frequencies $\Omega_1$ and $\Omega_2$ in Eq.~\eqref{excited_magnon}.
The time average of the integral over $k_z$ leads to
    \begin{align}
    N_{\rm static}\approx \sum_{\eta=\pm1}\left( \frac{|g_{\eta\kappa_1}|^2}{32|v_{\eta\kappa_1}|}\frac{1}{\alpha_G\Omega_1}+ \frac{|g_{\eta\kappa_2}|^2}{32|v_{\eta\kappa_2}|}\frac{1}{\alpha_G\Omega_2}\right),  
    \label{Nstatic}
\end{align}
in which $k_z=\eta\kappa_1$ ($k_z=\eta\kappa_2$) are roots of $\omega_{k_z}=\Omega_1$ ($\omega_{k_z}=\Omega_2$) and $v_\kappa={\partial \omega_{k_z}}/{\partial k_z}|_{k_z=\kappa}$ 
is the magnon group velocity. \( N_{\rm static}\)  measures the excitation efficiency that is inversely proportional to the broadening $\alpha_G \Omega$ and the group velocity $v_{\eta\kappa}$ of the excited magnons. 
Using Eq.~\eqref{coupling_constant},  $N_{\rm static}\propto e^{-2h|k_z|}(1-e^{-d|k_z|})^2/(dw)$ is exponentially suppressed by a large distance $h$ to the NV center at which is becomes inversely proportional to the wire cross section \(dw\), but depends non-monotonically on a small wire thickness $d$.

Substituting magnon mode amplitudes \eqref{excited_magnon} into the magnetization operator \eqref{magnetization_operator}, 
\begin{align}
    M_\beta(z>0)&=-\sqrt{2M_s\gamma\hbar}\sum_{k_z}\Big({\cal M}_{\beta}e^{ik_zz}{\beta}_{k_z}(t)+{\rm H.c.}\Big)\nonumber\\
    &=-\frac{\sqrt{2M_s\gamma\hbar}}{4v_{\kappa}}{\cal M}_{\beta}(\kappa)e^{i\kappa z}g_{\kappa}^*e^{-i\Omega_{1}t}\nonumber\\
    &+\frac{\sqrt{2M_s\gamma\hbar}}{4v_{\kappa'}}{\cal M}_{\beta}(\kappa')e^{i\kappa' z}g_{\kappa'}^*e^{-i\Omega_{2}t}+{\rm H.c}.,\nonumber\\
    M_\beta(z<0)&=-\frac{\sqrt{2M_s\gamma\hbar}}{4v_{-\kappa}}{\cal M}_{\beta}(-\kappa)e^{-i\kappa z}g_{-\kappa}^*e^{-i\Omega_{1}t}\nonumber\\
    &+\frac{\sqrt{2M_s\gamma\hbar}}{4v_{-\kappa'}}{\cal M}_{\beta}(-\kappa')e^{-i\kappa' z}g_{-\kappa'}^*e^{-i\Omega_{2}t}+{\rm H.c.},
\end{align}
in which $\kappa$ and $\kappa'$ with ${\rm Re}\,\kappa>0$ and ${\rm Re} \, \kappa'>0$ are governed by $\tilde{\omega}_{\kappa}-(\omega_{\rm NV}+|G_{eg}|)=0$ and $\tilde{\omega}_{\kappa'}-(\omega_{\rm NV}-|G_{eg}|)=0$.

\begin{figure}[htp!]
\centering
\includegraphics[width=0.45\linewidth]{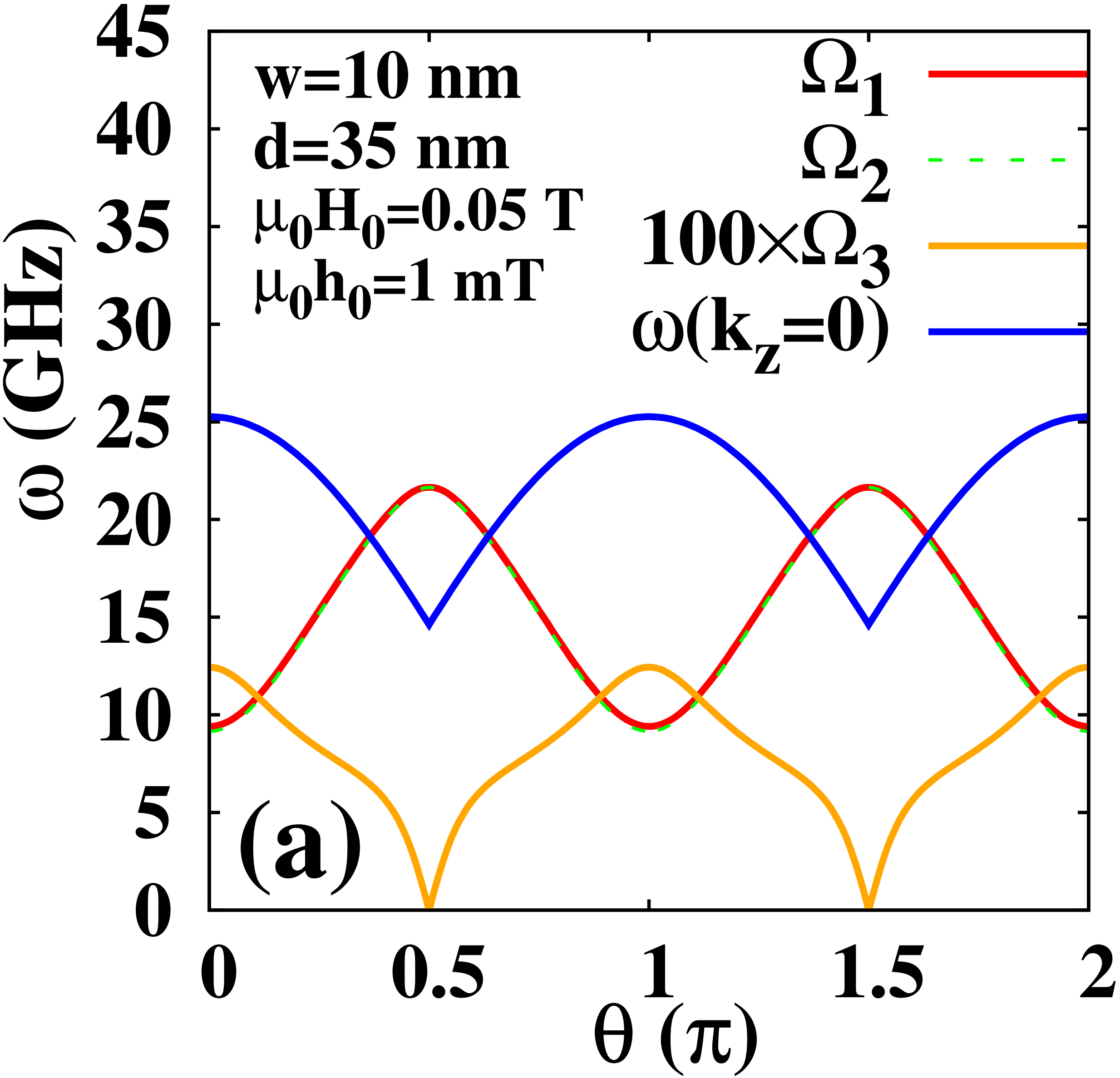}
\hspace{-0.23cm}
\includegraphics[width=0.55\linewidth]{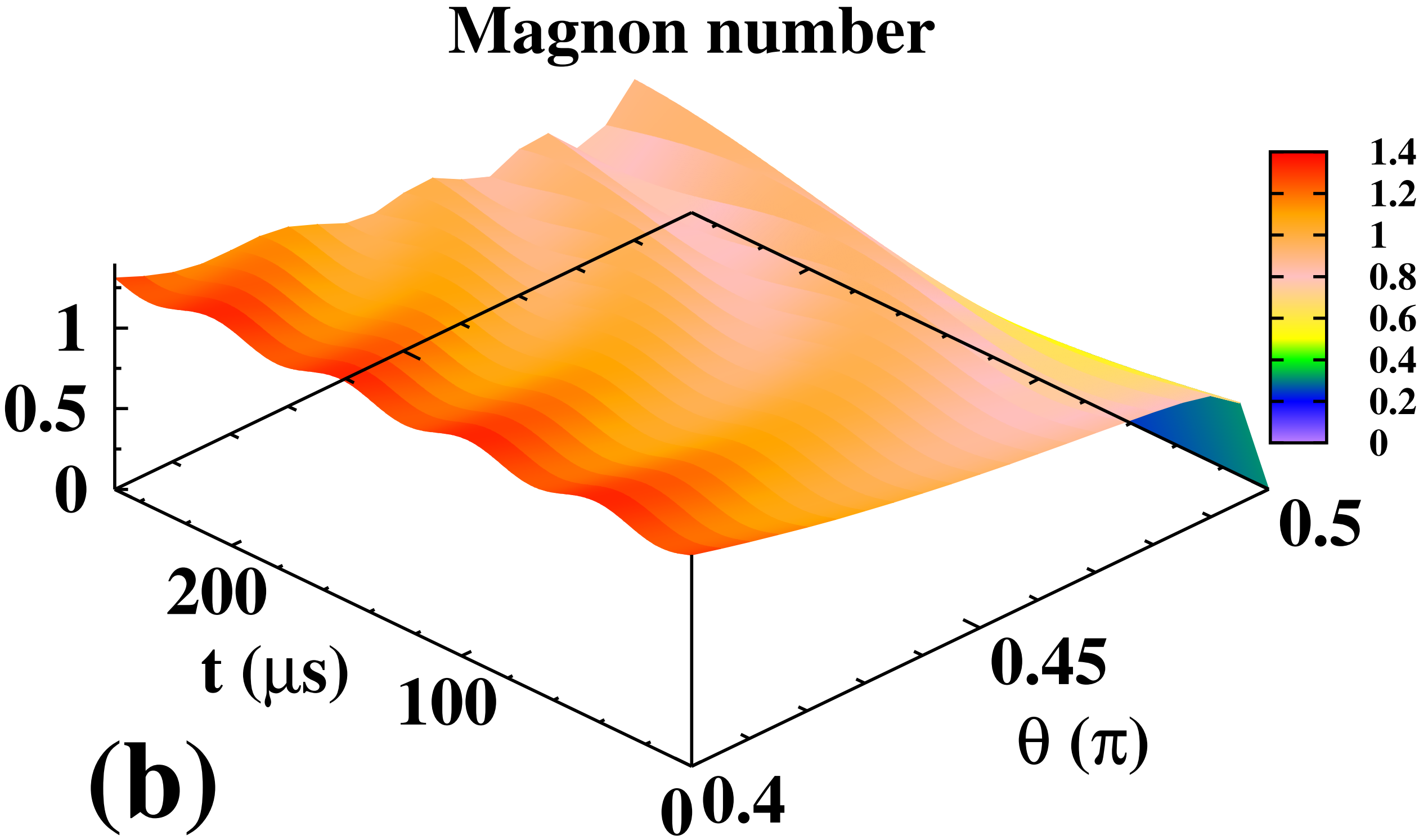}
\hspace{-0.2cm}
\includegraphics[width=0.46\linewidth]{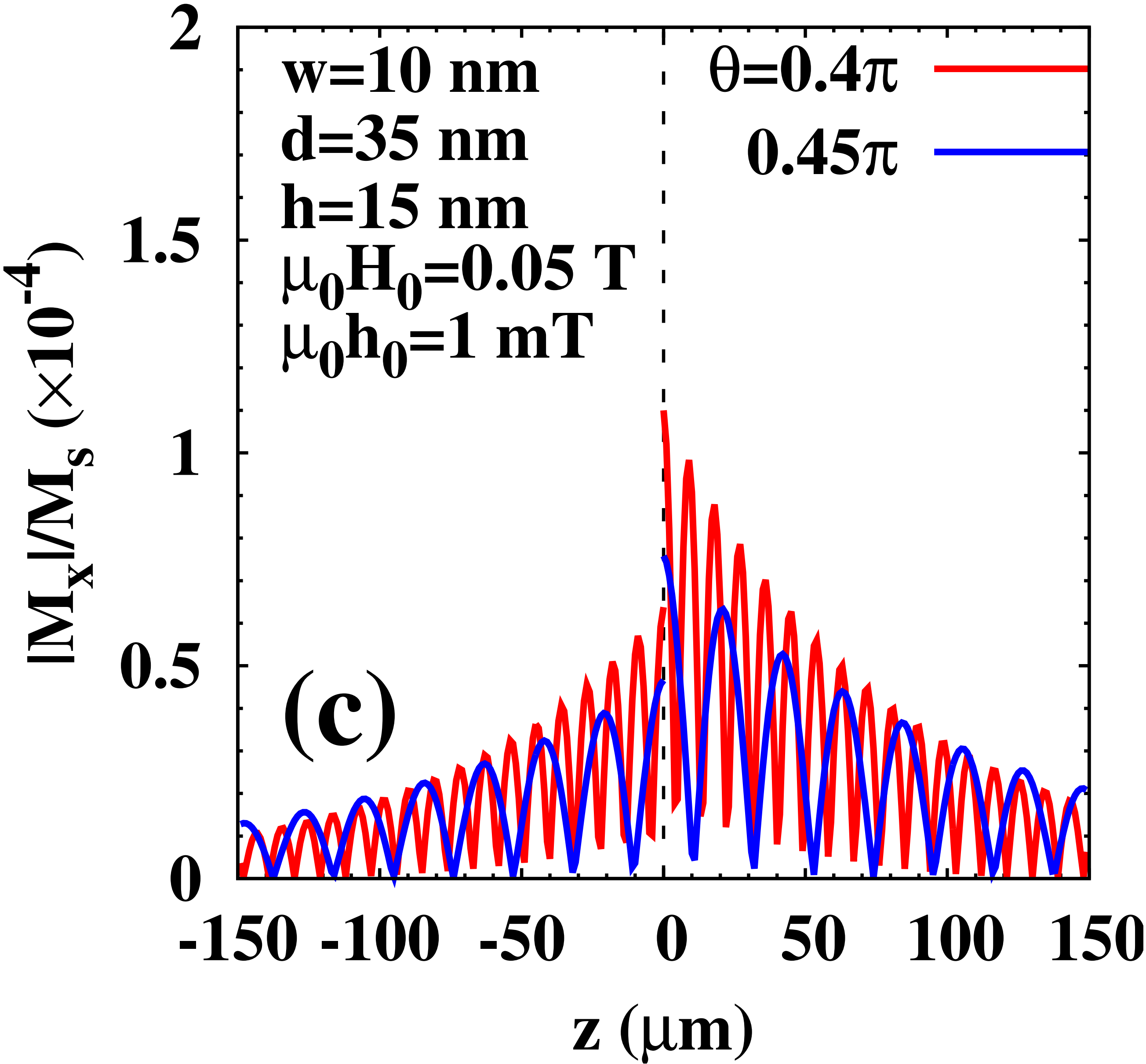}
\hspace{0.1cm}
\includegraphics[width=0.51\linewidth]{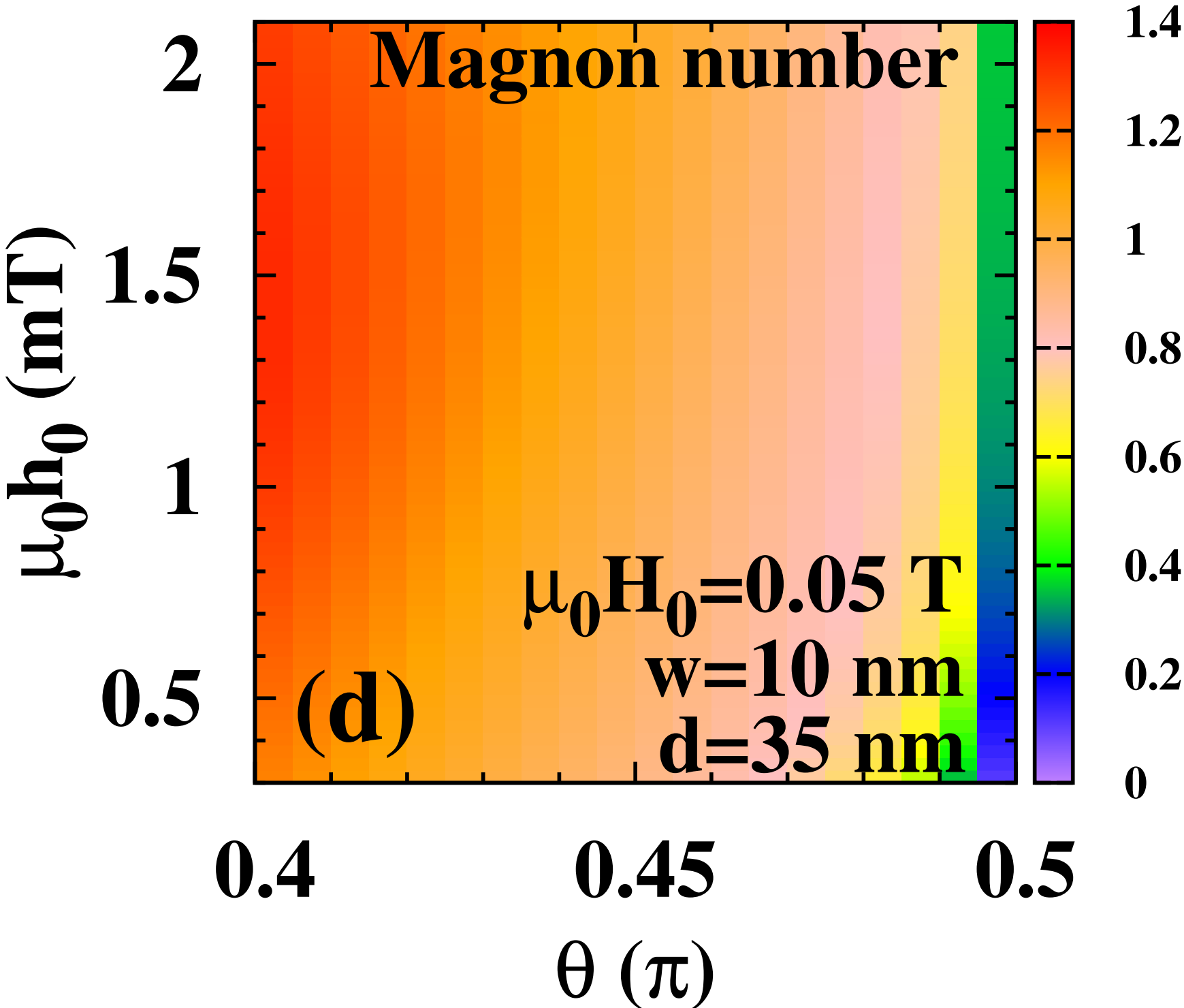}
\hspace{-0.23cm}
\caption{Magnon number distribution in a YIG nanowire (thickness $d=35$~nm, width $w=10$~nm) excited by the magnetic stray field emitted by a point-like NV-center $h=15$~nm above the wire at \(z=0\) that is actuated by an external microwave field $\mu_0h_0=1~{\rm mT}$ linearly polarized along $\hat{\bf y}$. 
A static magnetic field is applied at an angle $\theta$ with modulus  $\mu_0H_0=0.05~{\rm T}$. \textbf{(a)}:  Frequencies $\Omega_i$ of the NV-center stray field and the spin wave band minimum $\omega(k_z=0)$.  The excitation is resonant only close to the normal configuration  ${\bf H}_0\perp \hat{\bf z}$ \((\theta=\pi/2)\).
\textbf{(b)}: Total excited magnon number in the YIG nanowire in the resonant regime $\theta\in [0.4\pi,0.5\pi]$ showing a weak beating pattern. 
\textbf{(c)}: Weak spatial chirality of the resonantly excited magnetization distribution along the wire  $|M_x(z)|/M_s$ for $\theta=\{0.4\pi,0.45\pi\}$. \textbf{(d)}: Dependence of the excited magnon number on the microwave magnetic field $h_0$ and the direction of the static magnetic field $\theta$.}
\label{excitation1d}
\end{figure}

 As above, we consider a YIG nanowire of thickness $s=35~\mathrm{nm}$ and width $w=10$~nm biased by a weak applied magnetic field $\mu_{0}H_{0}=0.05$~T.  
In this resonant regime $\theta\in[0.4\pi,0.5\pi]$ and the parameters above, the average magnon number is around one except for $\theta\approx \pi/2$, as shown in Fig.~\ref{excitation1d}(b). 
The angle $\theta=\pi/2$ is special because the eigenstates of the NV center Eq.~\eqref{diagonalization} read
\begin{align}
    u_e&=\left(\begin{array}{c}
        \frac{\sqrt{2}}{2}, 0, \frac{\sqrt{2}}{2}
    \end{array}\right)^T,\nonumber\\
    u_g&=\left(\begin{array}{c}
        \frac{\omega_H}{({D_{\rm NV}^2}+4\omega_H^2-D_{\rm NV}\sqrt{D_{\rm NV}^2+4\omega_H^2})^{1/2}}  \\
        \frac{-\sqrt{2}i(D_{\rm NV}-\sqrt{D_{\rm NV}^2+4\omega_H^2})}{2({D_{\rm NV}^2}+4\omega_H^2-D_{\rm NV}\sqrt{D_{\rm NV}^2+4\omega_H^2})^{1/2}}  \\ -\frac{\omega_H}{({D_{\rm NV}^2}+4\omega_H^2-D_{\rm NV}\sqrt{D_{\rm NV}^2+4\omega_H^2})^{1/2}}
    \end{array}\right),
\end{align}
leading to $S_{eg}^y=u_e^{\dagger}\hat{S}^y_{\rm NV}u_g=0$ and $\Omega_1=\Omega_2$, such that the two terms in Eq.~\eqref{Number} cancel. The computed beating pattern of the magnon number caused by the interference of the two characteristic excitation frequencies $\{\Omega_1,\Omega_2\}$ in Fig.~\ref{excitation1d}(b) turns out to be relatively weak.

Figure~\ref{excitation1d}(c) illustrates the spatial distribution of the excited magnetization $|M_x(z)|/M_s$ for field directions $\theta \in \{0.4\pi,0.45\pi\}$ showing only a small chirality, which is expected when the equilibrium magnetization is almost parallel with the wire. The chirality becomes more pronounced at larger applied magnetic fields that significantly tilt the magnetization (not shown). 
Figure~\ref{excitation1d}(d) illustrates the dependence of the magnon number on the microwave magnetic field $h_0$ and the field angle $\theta$ is close to one.

\section{Derivation of master equation}

\label{derivation_master_equation}

The interaction Hamiltonian between the NV center and magnons in the subspace spanned by the NV-center states $\{|i,e\rangle,|i,g\rangle \}$ after the rotation-wave approximation reads 
\begin{align}
    {\hat  H}_{\rm int}&=\sum_{i=\{1,2\}}\sum_{{k}_z}\hbar g_{k_z}e^{i{k_z}\cdot{z}_{i}}\hat{\sigma}^{+}_{{\rm NV}_{i}}\hat{\beta}_{k_z}+{\rm H.c.},
\end{align}
where $g_{k_z}$ is the coupling constant, $\hat{\sigma}^{+}_{{\rm NV},{i}}=|i,e\rangle \langle i,g|$, and $\hat{\sigma}^{-}_{{\rm NV},{i}}=|i,g\rangle \langle i,e|$. 
The unperturbed Hamiltonians of NV centers $\hat{H}_{\rm{NV}}=\sum_{i=\{1,2\}}({\hbar \omega_{\rm NV}}/{2})\hat{\sigma}_{{\rm NV}_i}^z$ and magnons $\hat{\beta}_{k_z}$ in the magnetic nanowire $\hat{H}_{\rm {m}}=\sum_{k_z}\hbar\tilde{\omega}_{k_z}\hat{\beta}_{k_z}^{\dagger}\hat{\beta}_{k_z}$, 
where $\hat{\sigma}_{{\rm NV}_i}^z=|i,e\rangle \langle i,e|-|i,g\rangle \langle i,g|$, $\omega_{\rm NV}=\omega_e-\omega_g$ is the frequency of the two-level system of NV centers, and $\tilde{\omega}_{k_z}$ is the frequency of a magnon with wave number $k_z$. 
In the interaction picture, 
\begin{align}
    {\hat H}^{I}_{\rm int}(t)&=e^{i({\hat  H}_{m}+{\hat  H}_{\rm NV})t/\hbar}{\hat H}_{\rm int}e^{-i({\hat  H}_{m}+{\hat  H}_{\rm NV})t/\hbar}\nonumber\\
    &=\sum_{i=1,2}\sum_{{k}_z}\hbar g_{k_z}e^{i{k_z}{z}_{i}}\hat{\sigma}^{+}_{{\rm NV}_{i}}\hat{\beta}_{k_z}^{I}(t)e^{i\omega_{\rm NV}t}+{\rm H.c.},
\end{align}
where $\hat{\beta}_{k_z}^{I}(t)=e^{i{\hat  H}_{m}t/\hbar}\hat{\beta}_{k_z}e^{-i{\hat  H}_{m}t/\hbar}=e^{-i\tilde{\omega}_{k_z}t}\hat{\beta}_{k_z}$.
The density matrix $\hat{\rho}^I_{\rm SE}(t)$ of the entire system obeys the von Neumann equation in the interaction picture 
\begin{align}
    \frac{d\hat{\rho}^I_{\rm SE}(t)}{dt}&=-\frac{i}{\hbar}\left[\hat{H}^I_{\rm int}(t),\hat{\rho}^I_{\rm SE}(t)\right], 
\end{align}
which may be integrated and expanded to become
\begin{align}
    \hspace{-0.3cm}\hat{\rho}^I_{\rm SE}(t)&=\hat{\rho}^I_{\rm SE}(0)-\frac{i}{\hbar}\int_0^{t}ds\left[\hat{H}^I_{\rm int}(s),\hat{\rho}^I_{\rm SE}(s)\right]\nonumber\\
    \hspace{-0.3cm}&=\hat{\rho}^I_{\rm SE}(0)-\frac{i}{\hbar}\int_0^{t}ds\left[\hat{H}^I_{\rm int}(s),\hat{\rho}^I_{\rm SE}(0)\right]\nonumber\\
    &-\frac{1}{\hbar^2}\int_0^tds\int_0^sds'\left[\hat{H}^I_{\rm int}(s),[\hat{H}^I_{\rm int}(s'),\hat{\rho}^I_{\rm SE}(s')]\right].
\end{align}
By the time derivative, 
\begin{align}
    \frac{d\hat{\rho}^I_{\rm SE}(t)}{dt}&=-\frac{i}{\hbar}\left[\hat{H}^I_{\rm int}(t),\hat{\rho}^I_{\rm SE}(0)\right]\nonumber\\
    &-\frac{1}{\hbar^2}\int_0^tds\left[\hat{H}^I_{\rm int}(t),[\hat{H}^I_{\rm int}(s),\hat{\rho}^I_{\rm SE}(s)]\right].
\end{align}
Since the effective coupling with magnons $\Gamma_{ij}\ll \omega_{\rm NV}\sim \omega_m$ is weak, we may trace out the magnon bath ``\textit{E}" to obtain the equation of motion for the density matrix $\hat{\rho}^I$ of the subsystem ``\textit{S}'': 
\begin{align}
    \frac{d\hat{\rho}^I_{\rm S}(t)}{dt}&=-\frac{i}{\hbar}{\rm Tr}_E\left[\hat{H}^I_{\rm int}(t),\hat{\rho}^I_{\rm SE}(0)\right]\nonumber\\
    &-\frac{1}{\hbar^2}\int_0^tds{\rm Tr}_E\left[\hat{H}^I_{\rm int}(t),[\hat{H}^I_{\rm int}(s),\hat{\rho}^I_{\rm SE}(s)]\right]. 
\end{align}
Vice versa, owing to the weak coupling, the density matrix of the environment, i.e., here the magnon bath, is not significantly affected by the interaction, justifying the Born approximation
\begin{align}
    \hat{\rho}^I_{\rm SE}(s)&\approx\hat{\rho}^I(s)\otimes\hat{\rho}_E^{\rm eq}, 
\end{align}
where $\hat{\rho}_E^{\rm eq}$ is the equilibrium density matrix of the magnons.
Since $\hat{H}^I_{\rm int}(t)$ only contains the single-magnon operators $\hat{\beta}_{k_z}$ or $\hat{\beta}^{\dagger}_{k_z}$ and $\langle \hat{\beta}_{k_z}\rangle={\rm Tr}_E(\hat{\beta}_{k_z}\hat{\rho}_{E}^{\rm eq})=0$, $\langle \hat{\beta}^{\dagger}_{k_z}\rangle={\rm Tr}_E(\hat{\beta}^{\dagger}_{k_z}\hat{\rho}_{E}^{\rm eq})=0$, hence $-({i}/{\hbar}){\rm Tr}_E\left[\hat{H}^I_{\rm int}(t),\hat{\rho}^I_{\rm SE}(0)\right]=0$:
\begin{align}
    \frac{d\hat{\rho}^I_{\rm S}(t)}{dt}=-\frac{1}{\hbar^2}\int_0^tds{\rm Tr}_E\left[\hat{H}^I_{\rm int}(t),[\hat{H}^I_{\rm int}(s),\hat{\rho}^I_{\rm SE}(s)]\right]. 
\end{align}

The Markov approximation holds when the environment (magnon bath) returns rapidly to equilibrium irrespective of the perturbation, which requires a relaxation time that is short compared to the environment-induced system dynamics. This is the case in our NV-magnon coupled system with typical magnon relaxation times of $100$~ns that are much shorter than the magnon-induced NV center dynamics $1/\Gamma_{ii}\sim (300~{\rm kHz})\sim 3~{\rm \mu s}$.
Implementing the Markov approximation ${\rho}^I(s)\rightarrow{\rho}^I(t)$, changing the variable $s\rightarrow\tau \rightarrow t-\tau$, and extending the integration range $[0,t]\rightarrow[0,\infty]$, we arrive at the Lindblad master equation 
    \begin{align}
    \frac{d \hat{\rho}^{I}}{d t}&=-\frac{1}{\hbar^{2}}\int_{0}^{\infty}d\tau  ~{\rm Tr}_{E}\Big[{\hat H}^{I}_{\rm int}(t),[{\hat H}^{I}_{\rm int}(t-\tau),\hat{\rho}^{I}(t)\otimes \hat{\rho}^{\rm eq}_{E}] \Big].
    \label{Lindblad_master_equation}
\end{align}
Inserting our interaction Hamiltonian \eqref{interaction_Hamiltonian} leads to the integral 
\begin{align}
    &\int_{0}^{\infty}d\tau  ~{\rm Tr}_{E}\Big[{\hat H}^{I}_{\rm int}(t),[{\hat H}^{I}_{\rm int}(t-\tau),\hat{\rho}^{I}(t)\otimes \hat{\rho}^{\rm eq}_{E}] \Big]\nonumber\\
    &=\int_{0}^{\infty}d\tau \sum_{i=1,2}\sum_{j=1,2}\sum_{k_z}\Big[|g_{k_z}|^{2}\hbar^{2}e^{i\omega_{\rm NV}\tau}e^{i{k_z}({z}_{j}-{z}_{i})}\nonumber\\
    &\times\langle \hat{\beta}_{k_z}^{I}(\tau)\hat{\beta}_{k_z}^{I\dagger}\rangle(\hat{\sigma}^{+}_{{\rm NV},{j}}\hat{\sigma}^{-}_{{\rm NV},{i}}\hat{\rho}^{I}-\hat{\sigma}^{-}_{{\rm NV},{i}}\hat{\rho}^{I}\hat{\sigma}^{+}_{{\rm NV},{j}})\nonumber\\
    &+|g_{k_z}|^{2}\hbar^{2}e^{i\omega_{\rm NV}\tau}e^{i{k_z}({z}_{j}-{z}_{i})}\langle \hat{\beta}_{k_z}^{I^\dagger}\hat{\beta}_{k_z}^{I}(\tau)\rangle\nonumber\\
    &\times(\hat{\rho}^{I}\hat{\sigma}^{-}_{{\rm NV},{i}}\hat{\sigma}^{+}_{{\rm NV},{j}}-\hat{\sigma}^{+}_{{\rm NV},{j}}\hat{\rho}^{I}\hat{\sigma}^{-}_{{\rm NV},{i}})\nonumber\\
    &+|g_{k_z}|^{2}\hbar^{2}e^{-i\omega_{\rm NV}\tau}
    e^{i{k_z}({z}_{j}-{z}_{i})}\langle \hat{\beta}_{k_z}^{I^\dagger}(\tau)\hat{\beta}_{k_z}^{I}\rangle\nonumber\\
    &\times(\hat{\sigma}_{{\rm NV},i}^{-}\hat{\sigma}_{{\rm NV},j}^{+}\hat{\rho}^{I}-\hat{\sigma}_{{\rm NV},j}^{+}\hat{\rho}^{I}\hat{\sigma}_{{\rm NV},i}^{-})\nonumber\\&
    +|g_{k_z}|^{2}\hbar^{2}e^{-i\omega_{\rm NV}\tau}e^{i{k_z}({z}_{j}-{z}_{i})}\langle \hat{\beta}_{k_z}^{I}\hat{\beta}_{k_z}^{I\dagger}(\tau)\rangle\nonumber\\
    &\times(\hat{\rho}^{I}\hat{\sigma}_{{\rm NV},j}^{+}\hat{\sigma}_{{\rm NV},i}^{-}-\hat{\sigma}_{{\rm NV},i}^{-}\hat{\rho}^{I}\hat{\sigma}_{{\rm NV},j}^{+})\Big],
    \end{align}
where $\langle \hat{\beta}_{k_z}^{I\dagger}(t)\hat{\beta}_{k_z}^{I}(t-\tau)\rangle=\langle \hat{\beta}_{k_z}^{I\dagger}(\tau)\hat{\beta}_{k_z}^{I}\rangle={\rm Tr}[\hat{\beta}_{k_z}^{I\dagger}(\tau)\hat{\beta}_{k_z}^{I}\hat{\rho}^{\rm eq}_{E}]=e^{i\omega_{k_{z}}\tau} \langle\hat{n}_{k_{z}}\rangle$.
The master equation now becomes
\begin{align}
    &\frac{d \hat{\rho}^{I}}{d t}=-i\left[\sum_{\{i,j\}=1,2} \Omega_{ij}\hat{\sigma}_{{\rm NV},j}^{+}\hat{\sigma}_{{\rm NV},i}^{-},\hat{\rho}^{I}\right]+i\left[\Omega\sum_{i=1,2}\hat{\sigma}_{{\rm NV},i}^{z},\hat{\rho}^{I}\right]\nonumber\\
    &-\sum_{i,j}\Omega_{\downarrow, ij}\left(\frac{1}{2}\hat{\sigma}_{{\rm NV},j}^{+}\hat{\sigma}_{{\rm NV},i}^{-}\hat{\rho}^{I}+\frac{1}{2}\hat{\rho}^{I}\hat{\sigma}_{{\rm NV},j}^{+}\hat{\sigma}_{{\rm NV},i}^{-}-\hat{\sigma}_{{\rm NV},i}^{-}\hat{\rho}^{I}\hat{\sigma}_{{\rm NV},j}^{+}\right)\nonumber\\
    &-\sum_{i,j}\Omega_{\uparrow, ij}\left(\frac{1}{2}\hat{\sigma}_{{\rm NV},i}^{-}\hat{\sigma}_{{\rm NV},j}^{+}\hat{\rho}^{I}+\frac{1}{2}\hat{\rho}^{I}\hat{\sigma}_{{\rm NV},i}^{-}\hat{\sigma}_{{\rm NV},j}^{+}-\hat{\sigma}_{{\rm NV},j}^{+}\hat{\rho}^{I}\hat{\sigma}_{{\rm NV},i}^{-}\right),
    \label{master_equation}
\end{align}
with coefficients
\begin{align}
    \Omega^{\downarrow}_{ij}&=\sum_{k_z}2\pi|g_{k_z}|^{2}\delta({\omega}_{k_z}-\omega_{\rm NV})e^{i{k_z}({z}_{j}-{z}_{i})}( {n}(\hbar\omega_{\rm NV})+1)\nonumber\\
    &=\gamma_{ij}(\omega_{\rm NV})( {n}(\hbar\omega_{\rm NV})+1),\nonumber\\
\Omega^{\uparrow}_{ij}&=\sum_{k_z}2\pi|g_{k_z}|^{2}\delta({\omega}_{k_z}-\omega_{\rm NV})e^{i{k_z}({z}_{j}-{z}_{i})} {n}(\hbar\omega_{\rm NV})\nonumber\\
&=\gamma_{ij}(\omega_{\rm NV}) {n}(\hbar\omega_{\rm NV}),\nonumber\\
    \omega_{1, ij}&=\sum_{k_z}|g_{k_z}|^{2}e^{i{k_z}({z}_{j}-{z}_{i})}[1+ {n}(\hbar\omega_{\rm NV})]{\rm Re}\left(\frac{1}{\omega_{\rm NV}-\tilde{\omega}_{k_z}}\right),\nonumber\\
    \omega_{2, ij}&=-\sum_{k_z}|g_{k_z}|^{2}e^{i{k_z}({z}_{j}-{z}_{i})}{n}(\hbar\omega_{\rm NV}){\rm Re}\left(\frac{1}{\omega_{\rm NV}-\tilde{\omega}_{k_z}}\right),\nonumber\\
    \Omega_{ij}&=\omega_{1,ij}+\omega_{2, ij}=\sum_{k_z}|g_{k_z}|^{2}e^{i{k_z}({z}_{j}-{z}_{i})}{\rm Re}\left(\frac{1}{\omega_{\rm NV}-\tilde{\omega}_{k_z}}\right),\nonumber\\\Omega&=\omega_{2,ii}=-\sum_{k_z}|g_{k_z}|^{2}{n}(\hbar\omega_{\rm NV}){\rm Re}\left(\frac{1}{\omega_{\rm NV}-\tilde{\omega}_{k_z}}\right).
\end{align}

\end{appendix}

\end{document}